%% file: elliptix.tex
\RequirePackage{ifpdf}
\documentclass[11pt,a4paper]{JHEP3}
\usepackage{amsmath}
\usepackage{amsfonts}
\usepackage{slashed}
\usepackage{shuffle}
\usepackage{graphicx}

\usepackage[numbers,sort&compress]{natbib}

\newcommand{\eps}{\epsilon}
\newcommand{\ord}{\begin{cal}O\end{cal}}

\def\beq{\begin{equation}}
\def\eeq{\end{equation}}
\def\bsp#1\esp{\begin{split}#1\end{split}}

\newcommand{\cA}{\begin{cal}A\end{cal}}

\newcommand{\cE}{\begin{cal}E\end{cal}}

\newcommand{\cG}{\begin{cal}G\end{cal}}

\newcommand{\cR}{\begin{cal}R\end{cal}}

\newcommand{\cX}{\begin{cal}X\end{cal}}

\newcommand{\cZ}{\begin{cal}Z\end{cal}}
\newcommand{\EK}{{\rm K}}
\newcommand{\EE}{{\rm E}}
\newcommand{\EF}{{\rm F}}
\newcommand{\cra}{{\rm cr}}

\newcommand{\Efe}[4]{{\textrm{E}_4}\!\left(\begin{smallmatrix}#1\\#2\end{smallmatrix};#3,#4\right)}
\newcommand{\Ete}[4]{{\textrm{E}_3}\!\left(\begin{smallmatrix}#1\\#2\end{smallmatrix};#3,#4\right)}
\newcommand{\Ef}[3]{{\textrm{E}_4}\!\left(\begin{smallmatrix}#1\\#2\end{smallmatrix};#3\right)}
\newcommand{\Et}[3]{{\textrm{E}_3}\!\left(\begin{smallmatrix}#1\\#2\end{smallmatrix};#3\right)}
\newcommand{\ERegt}[2]{{\varepsilon_3}\!\left(\begin{smallmatrix}#1\\#2\end{smallmatrix}\right)}
\newcommand{\ERegf}[2]{{\varepsilon_4}\!\left(\begin{smallmatrix}#1\\#2\end{smallmatrix}\right)}
\newcommand{\gamt}[3]{{\widetilde{\Gamma}}\!\left(\begin{smallmatrix}#1\\#2\end{smallmatrix};#3\right)}
\newcommand{\gam}[3]{{\Gamma}\!\left(\begin{smallmatrix}#1\\#2\end{smallmatrix};#3\right)}

\title{Elliptic polylogarithms and iterated integrals on elliptic curves I: general formalism}

\author{Johannes Broedel$^a$\footnote{jbroedel@physik.hu-berlin.de} , 
Claude Duhr$^{b,c}$\footnote{claude.duhr@cern.ch} , 
Falko Dulat$^d$\footnote{dulatf@slac.stanford.edu} , 
Lorenzo Tancredi$^b$\footnote{lorenzo.tancredi@cern.ch}\; \\

{\it $^a$Institut f\"{u}r Mathematik und Institut f\"{u}r Physik,
Humboldt-Universit\"{a}t zu Berlin\\
\phantom{$^a$}IRIS Adlershof, Zum Grossen Windkanal 6, 12489 Berlin, Germany} \\
{\it $^b$Theoretical Physics Department, CERN, Geneva, Switzerland} \\
{\it$^c$Center for Cosmology, Particle Physics and Phenomenology (CP3),\\
\phantom{$^c$}Universit\'e Catholique de Louvain, 1348 Louvain-La-Neuve, Belgium}\\
{\it $^d$SLAC National Accelerator Laboratory, Stanford University, Stanford, CA 94309, USA } 

}

\preprint{CERN-TH-2017-273, CP3-17-57,  HU-EP-17/29,
HU-Mathematik-2017-09, SLAC-PUB-17194}

\abstract{
We introduce a class of iterated integrals, defined through a set of linearly 
independent integration kernels on elliptic curves. As a direct generalisation
of multiple polylogarithms, we construct our set of
integration kernels ensuring that they have at most simple poles, implying that the iterated integrals have at most logarithmic singularities. We
study the properties of our iterated integrals and their relationship to the {multiple elliptic polylogarithms}
from the mathematics literature. On the one hand, we find that our iterated integrals span essentially
the same space of functions as the multiple elliptic polylogarithms.
On the other, our formulation allows for a more direct use 
to solve a large variety of problems in high-energy physics. We demonstrate 
the use of our functions in the evaluation
of the Laurent expansion of some hypergeometric functions for values of the
indices close to half integers.
}

\begin{document}

\catcode`\@=11
\font\manfnt=manfnt
\def\Watchout{\@ifnextchar [{\W@tchout}{\W@tchout[1]}}
\def\W@tchout[#1]{{\manfnt\@tempcnta#1\relax%
  \@whilenum\@tempcnta>\z@\do{%
    \char"7F\hskip 0.3em\advance\@tempcnta\m@ne}}}
\let\foo\W@tchout
\def\dubious{\@ifnextchar[{\@dubious}{\@dubious[1]}}
\let\enddubious\endlist
\def\@dubious[#1]{%
  \setbox\@tempboxa\hbox{\@W@tchout#1}
  \@tempdima\wd\@tempboxa
  \list{}{\leftmargin\@tempdima}\item[\hbox to 0pt{\hss\@W@tchout#1}]}
\def\@W@tchout#1{\W@tchout[#1]}
\catcode`\@=12

\input{intro}

\input{genus0}

\input{summary}

\input{torus}

\input{weierstrass_curve}

\input{algorithm}

\input{general}

\input{examples}

\input{conclusion}

\section*{Acknowledgements}
The authors are grateful to Nils Matthes, Erik Panzer, Brenda Penante and Ettore Remiddi for stimulating discussions, and to the ETH Z\"urich and the Pauli Center for Theoretical Studies for the organisation of the workshop ``The elliptic/missing Feynman integrals'', where some of the ideas presented in this paper were first discussed.
This research was supported by the the ERC grant 637019 ``MathAm'',
the U.S. Department of Energy (DOE) under contract DE-AC02-76SF00515
and the Munich Institute for Astro- and Particle Physics (MIAPP) of the DFG cluster of excellence ``Origin and Structure of the Universe''.

\appendix 

\input{regularisation}

\bibliography{bib}

\end{document}

%% file: intro.tex

\section{Introduction}
\label{sec:intro}

The discovery of the Higgs boson at the LHC and the absence of clear signs of
New Physics close to the electroweak scale provide a strong confirmation of the
Standard Model of particles physics (SM) up to the TeV scale.  At the same
time, the SM fails spectacularly at explaining a wide spectrum of important
phenomena, like the overwhelming experimental indications for the existence of
Dark Matter, and also at providing a consistent framework to describe all known
fundamental interactions, including gravity.  Given the increasing precision of
the measurements carried out at the LHC, equally precise calculations in the SM
become then an essential tool to uncover shortcomings of the SM and discover
possible elusive signs of New Physics.

The precision calculation of physical observables in perturbative quantum field
theories relies on the computation of Feynman graphs with many external legs
and loops.  As a result of more than a decade of investigation, a lot has been
understood on the mathematical properties of Feynman integrals. These
developments, in turn, have led to powerful techniques for the analytical and
numerical calculation of loop integrals.  In particular, the discovery that a
huge class of Feynman graphs can be naturally expressed in terms of multiple
polylogarithms \cite{Goncharov:1995,Remiddi:1999ew,
Goncharov:2001,Vollinga:2004sn}, has made possible the calculation of
higher-order corrections to a large number of processes of crucial importance
for the precision physics program carried out at the LHC, but that were
previously thought to be out of reach.  In spite of having been known to
mathematicians for a very long time~\cite{Nielsen,Kummer}, during the last
decade multiple polylogarithms have become again an active field of research,
as the discovery of their underlying Hopf algebra structure has paved the way
towards a thorough understanding of their analytical and numerical properties
\cite{Goncharov:2010jf,Ablinger:2011te,Duhr:2012fh}, in particular allowing to
systematize the study of non-trivial functional identities between them.

While multiple polylogarithms have a large range of applicability, they are not
the end of the story.  Recently, a growing number of integrals appearing in
higher-order calculations in the SM, but also in ${\cal N}=4$ super-Yang--Mills
theory or open string theory, have been identified, whose analytical
calculation requires functions beyond multiple polylogarithms
\cite{Aglietti:2007as,CaronHuot:2012ab,Bloch:2016izu,Remiddi:2013joa,Laporta:2004rb,Bloch:2013tra,Adams:2013nia,Adams:2014vja,Adams:2015gva,Adams:2015ydq,Remiddi:2016gno,Adams:2016xah,Bonciani:2016qxi,vonManteuffel:2017hms,Primo:2017ipr,Ablinger:2017bjx,Chen:2017pyi,Bourjaily:2017bsb,Chen:2017soz,Broedel:2014vla}.
Indeed, the appearance of non-polylogarithmic structures in quantum field
theory had been noticed already more than fifty years ago in the computation of
the two-loop corrections to the electron self-energy in QED~\cite{Sabry}.  The
origin of this new class of functions could be traced back to a particular
Feynman graph, the so-called two-loop massive sunrise graph, whose analytical
properties have been extensively analyzed from different perspectives in the
physics and mathematics
literature~\cite{Broadhurst:1987ei,Bauberger:1994by,Bauberger:1994hx,Laporta:2004rb,Bloch:2013tra,Adams:2013nia,Remiddi:2013joa,Adams:2014vja,Adams:2015gva,Adams:2015ydq,Remiddi:2016gno,Adams:2016xah,Remiddi:2017har,Hidding:2017jkk}.\footnote{A
  proposal for the numerical evaluation of the functions which appear in the
calculation of the two-loop massive sunrise graph has been recently put forward
in ref.~\cite{Passarino:2016zcd}.} While many analytical representations for
the sunrise integral have been worked out, a way to generalize them to more
complicated Feynman integrals remains matter of debate.

From a more mathematical point of view, multiple polylogarithms are obtained by
integrating rational functions on a Riemann surface of genus zero -- a Riemann
sphere.  This perspective suggests as natural generalisation of multiple
polylogarithms the class of functions obtained by integrating rational
functions on Riemann surfaces of higher genus.  Indeed, it is by now very well
known that the two-loop massive sunrise graph, together with many other
examples from quantum field theory and string theory, can be represented by
functions on a surface of genus one -- a torus or, equivalently, an elliptic
curve.  Genus-one generalisations of multiple polylogarithms have been studied
by mathematicians~\cite{BeilinsonLevin, LevinRacinet, BrownLevin} and, quite
naturally, have been dubbed {\it multiple elliptic polylogarithms}. Many
important properties of these functions are by now well understood; in
particular, a very important result of~\cite{BrownLevin} was to introduce a set
of integration kernels defined on the torus with at most simple poles, and to
show that they allow to perform the integration of any rational function on the
corresponding elliptic curve, e.g.~the space of functions is closed under
taking primitives.

In spite of these appealing results, the use multiple elliptic polylogarithms
in the calculation of physically relevant Feynman integrals has been impeded
and complicated by the fact that most of the mathematical literature makes only
use of the torus formulation of elliptic curves.  While in that formulation the
integration kernels used to define elliptic polylogarithms have a very clean
geometrical interpretation, their application to solve problems in high energy
physics is not straightforward and for practical purposes a different language
appears to be preferable.  The main goal of this paper is therefore to provide
a different formulation of elliptic polylogarithms in terms of simple
integration kernels, which will hopefully turn out to be easier to use in the
context of high energy physics calculation.  As a fundamental result of this
paper, we will show that, when our set of integration kernels is mapped to the
torus, then they (essentially) span the same space as defined by the multiple
elliptic polylogarithms found in the mathematical literature.  Of course, we
find that multiple polylogarithms are just a subset of multiple elliptic
polylogarithms.

While in this paper we will focus mainly on the mathematical details at the
basis of the construction of elliptic polylogarithms, we will also provide two
examples to  demonstrate the flexibility of our setup and its applicability to
a variety of different problems.  An interesting and very non-trivial
mathematical problem is the computation of the Laurent expansion of different
kinds of hypergeometric functions, when the small expansion parameter (say
$\epsilon$) appears in their indices. In particular, it is well known that if
the indices are expanded around half-integer values, at order zero in the
expansion one is in general left with complete elliptic integrals of first and
second kind.  In this case, however, no closed analytical formula in terms of
familiar functions is known for the higher-order Laurent coefficients.  As a
natural application of our formalism, we will show how both in the case of the
Gauss' hypergeometric function ${}_2F_1$ and of the Appel function $F_1$, the
coefficients of their Laurent expansion around half-integer values can easily
and algorithmically be expressed in terms of our elliptic polylogarithms.  The
computations presented here are mainly of mathematical interest.  The most
prominent example in high-energy physics, the sunrise integral, will be
calculated in our elliptic language in a companion paper \cite{plumber_paper},
which we provide as a hands-on guide to the use of our functions in high-energy
physics for those who prefer not to have to dwell into all mathematical details
reported here.

The outline of the paper is as follows: after reviewing multiple polylogarithms
as the result of integrating rational functions on the Riemann sphere in
section~\ref{sec:genus0}, we present the main result of the paper in section
\ref{sec:summary}: a class of multiple elliptic polylogarithms. In
\mbox{section~\ref{sec:torus}} we describe iterated integrals in the torus
formulation of elliptic curves and discuss the relation between the torus and
our formulation in section \ref{sec:torus_to_Weierstrass}. A general algorithm
for solving integrals with roots of cubic polynomials based on our formalism is
provided in section \ref{sec:algorithm}.  In section \ref{sec:general} we
extend our formulation to elliptic curves defined by a general quartic
polynomial, and in section \ref{sec:examples} we use our framwork to compute
several hypergeometric functions ${}_2F_1$ and Appell $F_1$ functions. Finally,
we draw our conclusions in section \ref{sec:conclusion}.  In
appendix~\ref{app:shuffle_regularisation} we discuss some technical aspects
about regularisation of iterated integrals omitted in the main text.

%% file: genus0.tex

\section{Integrating rational functions on the Riemann sphere}
\label{sec:genus0}

In this section we give a short review of multiple polylogarithms and how they arise when integrating rational functions on the Riemann sphere. The material in this section is well known, but we discuss it in detail because it serves as a motivation and an illustration of the ideas that will be introduced in subsequent sections.

Let $R(x)$ be a rational function on the Riemann sphere $\mathbb{CP}^1$ with poles at $x=c_i$, for some complex numbers $c_i$. Our goal is to compute the integral of $R(x)$ along some path on the Riemann sphere. The precise form of the path is immaterial for our purposes, and we are only interested in computing a primitive of $R$. The value of any definite integral along some path can then be obtained by comparing the value of the primitive at the endpoints of the path (plus taking into account possible windings around poles).

Using partial fractioning, we can reduce the computation of the integral to a linear combination of elementary integrals\footnote{Since we are only interested in primitives, we could of course add any constant and still obtain a valid primitive.},
\beq
\int \frac{dx}{x^k} = \frac{x^{1-k}}{1-k} \textrm{~~and~~}\int \frac{dx}{(x-c_i)^k} = \frac{(x-c_i)^{1-k}}{1-k}\,.
\eeq
The previous relations are only valid for $k\neq1$. Indeed, even if we start from a rational function $R$, it is not always possible to find a \emph{rational} primitive, but we need to enlarge our space of functions. In this particular case, the obstruction to finding a rational primitive is related to the existence of the logarithm,
\beq\label{eq:genus0_simple_pole}
\int \frac{dx}{x} = \log x \textrm{~~and~~}\int \frac{dx}{x-c_i} =\log\left(1-\frac{x}{c_i}\right)\,.
\eeq
The reason for the appearance of an obstruction to finding a rational primitive is tightly linked to the fact that the integrands in eq.~\eqref{eq:genus0_simple_pole} have simple poles. Indeed, the residue theorem implies that the integral along a closed curve encircling a simple pole does not vanish, and as a consequence the integral defines a multi-valued function. Since rational functions are always single-valued, the integral of a function with a simple pole can in general not be expressed in terms of rational functions alone.

Next, we want to extend this analysis to iterated integrals over rational functions on the Riemann sphere. We then need to consider new kinds of obstructions, which arise from integrals that cannot be expressed in terms of rational functions and logarithms alone. Here we consider the simplest obstructions that one can encounter when studying iterated integrals on the Riemann sphere, \emph{multiple polylogarithms} (MPLs), which are defined recursively by~\cite{Goncharov:2001}
\beq\label{eq:MPL_def}
G(c_1,\ldots,c_n;x) = \int_{0}^xdt\,f(c_1,t)\,G(c_2,\ldots,c_n;t)\,,
\eeq
with
\beq \label{eq:MPL_kernel}
f(c,t) = \frac{1}{t-c}\,.
\eeq
The recursion starts at $G(;x)=1$. We assume that the $c_i$ are independent of $x$. In the case where $(c_1,\ldots,c_n)=(0,\ldots,0)$ the integral in eq.~\eqref{eq:MPL_def} is divergent, and we define instead
\beq\label{eq:log_def}
G(\underbrace{0,\ldots,0}_{n\textrm{ times}};x) = \frac{1}{n!}\log^nx\,,\qquad \log x=\int_1^x\frac{dt}{t}\,.
\eeq
MPLs satisfy some well-known properties, which we now quickly review. First, MPLs form a shuffle algebra, which allows one to express the product of two MPLs as a linear combination of the same class of functions,
\beq\label{eq:shuffle_G}
G(\vec c_1;x)\,G(\vec c_2;x) = \sum_{\vec c=\vec c_1\shuffle \vec c_2}G(\vec c;x)\,,
\eeq
where the sum runs over all shuffles of $\vec c_1$ and $\vec c_1$, i.e., over all permutations of $\vec c_1 \cup \vec c_2$ that preserve the relative orderings within $\vec c_1$ and $\vec c_2$. 
Second, if we consider the vector space of all MPLs with coefficients that are rational functions, then this space is closed under taking primitives. More precisely, if $\cR$ denotes the field of rational functions with poles at most at points $c_i\in S$, let $\cA_{\textrm{MPL}}$ denote the $\cR$-algebra generated by all MPLs with singularities at most for $x=c_i$. $\cA_{\textrm{MPL}}$ is graded by the weight (the multiplication is given by the shuffle product in eq.~\eqref{eq:shuffle_G}),
\beq
\cA_{\textrm{MPL}} = \bigoplus_{k=0}^\infty \cA_{\textrm{MPL},k}\,,\qquad \cA_{\textrm{MPL},k_1}\cdot \cA_{\textrm{MPL},k_2}\subset \cA_{\textrm{MPL},k_1+k_2}\,,
\eeq
with 
\beq
\cA_{\textrm{MPL},k} = \left\langle G(c_1,\ldots,c_k;x) : c_i\in S\right\rangle_{\cR}\,.
\eeq
$\cA_{\textrm{MPL}}$ is closed under both differentiation and integration.
In particular, for every \mbox{$f\in \cA_{\textrm{MPL}}$}, there is a primitive $F\in \cA_{\textrm{MPL}}$ such that $\partial_xF=f$.

The goal of this paper is to explain how MPLs and their properties can be generalised to elliptic
curves. We define a class of iterated integrals with logarithmic singularities that serve as a basis
for  obstructions that one can encounter when integrating a rational function on an elliptic curve,
and we show that their properties are very similar to the properties of MPLs.

%% file: summary.tex

\section{A class of elliptic polylogarithms}
\label{sec:summary}
The goal of this section is to present the main result of this paper: a generalisation of multiple polylogarithms to elliptic curves. The content of this section is self-contained and provides a summary of our main results. In addition, we present a concise review of the minimal background on elliptic curves needed to understand the construction of elliptic polylogarithms. For a more thorough review of the mathematical background on elliptic curves and for proofs of the results, we refer to subsequent sections and to the mathematical literature (see, e.g., ref.~\cite{silverman}).

\subsection{A lightning review of elliptic curves and elliptic integrals}
\label{sec:plumber_review}
Consider a cubic polynomial of the form 
\beq
P_3(x) = (x-a_1)(x-a_2)(x-a_3)\,.
\eeq
For concreteness, we assume that the $a_i$ are all distinct and real, and they are ordered according to $a_1<a_2<a_3$. Such a polynomial defines an \emph{elliptic curve} $\cE$ as the solution set of the polynomial equation 
\beq\label{eq:cubic_eq}
y^2=P_3(x) = (x-a_1)(x-a_2)(x-a_3)\,.
\eeq
Throughout this section we only consider elliptic curves defined by a cubic polynomial, and we will discuss elliptic curves defined by a quartic polynomial in section~\ref{sec:general}.

In the following it is convenient to work in projective space $\mathbb{CP}^2$, and we interpret the
polynomial equation in terms of homogeneous coordinates $[x,y,1]$. Seen as a curve in projective
space, $\cE$ also contains the point at infinity $[0,1,0]$. The points $[a_i,0,1]$ always lie on the
elliptic curve. In the following we will refer to the points $[0,1,0]$ and $[a_i,0,1]$ as
\emph{branch points}  (by abuse of language, we will often refer to the $a_i$ and $a_4\equiv\infty$
themselves as branch points). In addition, for every $x\neq a_i$, there are precisely two points
\mbox{$[x,\pm y,1]$ on $\cE$.}

An elliptic curve defines a compact Riemann surface of genus one. It is natural to ask what is the appropriate generalisation of a rational function to an elliptic curve. A \emph{rational function on the elliptic curve} $\cE$ is defined to be a rational function in the two variables $(x,y)$, subject to the constraint $y^2=P_3(x)$. In other words, a rational function $R$ on $\cE$ is an expression of the form
\beq
R(x,y) = \frac{p_1(x) + p_2(x)\,y}{q_1(x) + q_2(x)\,y} = \frac{p_1(x) + p_2(x)\,\sqrt{P_3(x)}}{q_1(x) + q_2(x)\,\sqrt{P_3(x)}}\,,
\eeq
where $p_i$ and $q_i$ are polynomials in $x$. If we multiply the numerator and the denominator 
by the conjugate of the denominator, $q_1(x) - q_2(x)\,\sqrt{P_3(x)}$, then we can write $R$ in the alternative form
\beq\label{eq:R_decomp}
R(x,y) = R_1(x) + \frac{1}{y}\,R_2(x) = R_1(x) + \frac{1}{\sqrt{P_3(x)}}\,R_2(x)\,,
\eeq
for some rational functions $R_i$. At this point we have to make a comment about the choice of the branch of the square root. In the case where the roots of $P_3$ are real and ordered according to $a_1<a_2<a_3$, we let
\begin{align}
\nonumber\sqrt{P_3(x)}&\,\equiv\sqrt{|P_3(x)|}\,\left[-i\,\theta(x\le a_1) + \theta(a_1<x\le a_2)+i\,\theta(a_2<x\le a_3) - \theta(a_3<x)\right]\\
&\,=\sqrt{|P_3(x)|}\times\left\{\begin{array}{ll}
-i\,,& x\le a_1\,,\\
\phantom{-}1\,,& a_1<x\le a_2\,,\\
\phantom{-}i\,,& a_2<x\le a_3\,,\\
-1\,,& a_3<x\,.
\end{array}\right.
\label{eq:root3}
\end{align}

Our goal is to study generalisations of polylogarithms to elliptic curves. We start by discussing the simpler case of a single integral of a rational function on $\cE$. Such integrals have been studied extensively in mathematics during the 19$^{\textrm{th}}$ century under the name of \emph{elliptic integrals}. We briefly review the computation of elliptic integrals, as some of the concepts will prove useful to define iterated integrals on elliptic curves. Using the decomposition in eq.~\eqref{eq:R_decomp}, we see that the contribution from $R_1(x)$ is an ordinary integral of a rational function and can be performed in terms of rational functions and logarithms. We will therefore focus on the contribution from $R_2(x)$. After partial fractioning, we only need to consider integrals of the form
\beq
\int \frac{dx}{y}\,x^k \textrm{~~and~~} \int\frac{dx}{y\,(x-c)^k}\,,
\eeq
where $k$ is an integer and $c$ is a constant. Using integration by parts, we can reduce every integral of this type to a linear combination of the following integrals:
\beq\label{eq:elliptic_integrals}
\int \frac{dx}{y}\,, \qquad \int \frac{x\,dx}{y}\,,\qquad \int\frac{dx}{y\,(x-c)}\,.
\eeq
These integrals cannot be simplified further, and should be thought of as the analogues of the concept of `master integrals' familiar from the the physics literature. Via a judicious change of variables, each of these integrals can be evaluated in terms of (incomplete) elliptic integrals of the first, second and third kind\footnote{In the literature, these integrals usually depend on an angular variable $x=\sin\varphi$.}
\beq\bsp\label{eq:FEPi}
\textrm{F}(x|m^2) &\,= \int_0^x\frac{dt}{\sqrt{(1-t^2)(1-m^2t^2)}}\,,\\
\textrm{E}(x|m^2) &\,= \int_0^xdt\,\frac{1-m^2t^2}{\sqrt{(1-t^2)(1-m^2t^2)}}\,,\\
\Pi(n^2,x|m^2) &\,= \int_0^x\frac{dt}{\sqrt{(1-t^2)(1-m^2t^2)}}\,\frac{1}{1-n^2t^2}\,.
\esp\eeq
The elliptic integrals in eq.~\eqref{eq:elliptic_integrals}, or equivalently in eq.~\eqref{eq:FEPi}, provide elementary obstructions to finding a rational primitive when integrating on an elliptic curve. This is in complete analogy with the role played by the logarithm in the case of the Riemann sphere, and we will see at the end of this section that, in a certain sense, we can indeed identify the incomplete elliptic integral of the third kind $\Pi$ with an elliptic generalisation of the logarithm.

The elliptic integrals in eq.~\eqref{eq:elliptic_integrals} allow one to define certain `invariants' that are attached to an elliptic curve. The \emph{periods} of $\cE$ are defined by integrating $dx/y$ between two branch points. We define
\beq\bsp
\omega_1 &\,\equiv 2c_3\int_{a_1}^{a_2}\frac{dx}{y} = 2\, \EK(\lambda) \,,\\
\omega_2 &\,\equiv 2c_3\int_{a_3}^{a_2}\frac{dx}{y}= i\, 2\, \EK(1-\lambda)\,,
\esp\eeq
with 
\beq\label{eq:c3_def}
c_3=\frac{\sqrt{a_{31}}}{2}\,,\qquad a_{ij} =a_i-a_j\,,\qquad \lambda \equiv \frac{a_{21}}{a_{31}}\,,
\eeq
and $\textrm{K}$ denotes the complete elliptic integral of the first kind, $\textrm{K}(\lambda) = \textrm{F}(1|\lambda)$.
If $\cE$ is non-degenerate (in particular if the roots of $P_3$ are distinct), then the two periods are linearly independent over the real numbers. Moreover,
all other integrals of this type evaluate to integer linear combinations of the $\omega_i$,
\beq
2c_3\int_{a_{i}}^{a_j}\frac{dx}{y} = m_{ij}\,\omega_1 + n_{ij}\,\omega_2\,,\qquad m_{ij}, n_{ij}\in \mathbb{Z}\,.
\eeq
Note that when the roots are real, $\omega_1$ can be chosen as real and positive, and $\omega_2$ has a positive imaginary part.

Similarly, the \emph{quasi-periods} of $\cE$ are defined by
\beq\bsp\label{eq:quasi-periods_def}
\eta_1&\,=\frac{1}{4}\int_{a_1}^{a_2}dx\,\widetilde{\Phi}_3(x) =
\textrm{E}(\lambda) - \frac{2-\lambda}{3}  \, \textrm{K}(\lambda) \,,\\
\eta_2 &\,= \frac{1}{4}\int_{a_3}^{a_2} dx\, \widetilde{\Phi}_3(x)=  
- i\,  \textrm{E}(1-\lambda) +i\, \frac{1+\lambda}{3}  \, \textrm{K}(1-\lambda) \,,
\esp\eeq
with $\textrm{E}(\lambda) = \textrm{E}(1|\lambda)$ and we defined
\beq\label{eq:Phi_3_tilde}
\widetilde{\Phi}_3(x,\vec a) = \frac{1}{c_3\,y} \left( -x +\frac{s_1(\vec a)}{3} \right) \,,
\eeq
where $s_n(\vec a)\equiv s_n(a_1,a_2,a_3)$ is the elementary symmetric polynomial of degree $n$ in three variables.
Just like in the case of the periods, a different choice of the branch points in the integration limits will result in an integer linear combination of $\eta_1$ and $\eta_2$. The periods and quasi-periods are not independent, but they are related through the \emph{Legendre relation},
\beq\label{eq:Legendre}
\omega_1\,\eta_2- \omega_2\,\eta_1 = -i\pi\,.
\eeq

The periods and quasi-periods are strictly speaking not invariants of $\cE$, but there may be different values of $\omega_i$ and $\eta_i$ that correspond to the same elliptic curve (for example, we can perform a global rescaling of the periods without changing the geometry). There is an invariant, called the \emph{j-invariant}, that uniquely characterises an elliptic curve,
\beq\label{eq:j-invariant}
j = 256\,\frac{(1-\lambda(1-\lambda))^3}{\lambda^2(1-\lambda)^2}\,.
\eeq
Two elliptic curves that have the same $j$-invariant are isomorphic.

\subsection{A class of elliptic polylogarithms}
We now introduce a class of elliptic generalisations of MPLs,
\beq\label{eq:eMPLs_def}
\Ete{n_1 & \ldots & n_k}{c_1 & \ldots& c_k}{x}{\vec a} = \int_0^xdt\,\varphi_{n_1}(c_1,t,\vec a)\,\Ete{n_2 & \ldots & n_k}{c_2 & \ldots& c_k}{t}{\vec a}\,,
\eeq
with $n_i\in\mathbb{Z}$ and $c_i\in\widehat{\mathbb{C}}\equiv\mathbb{C}\cup\{\infty\}$, and the recursion starts with $\textrm{E}_3(;x,\vec a)=1$. The vector $\vec a=(a_1,a_2,a_3)$  encodes the zeroes of the polynomial $P_3$, and therefore defines the elliptic curve $\cE$. In the following we will always assume that the elliptic curve is fixed, and we will keep the dependence of all quantities on $\vec a$ implicit. The integration kernels $\varphi_{n}$ that appear in eq.~\eqref{eq:eMPLs_def} are chosen such as to satisfy the following basic properties:
\begin{enumerate}
\item The functions $\varphi_{n}$ are non-trivial, i.e., they cannot be written as total derivatives of a rational function on $\cE$ (because otherwise the  integration would be trivial). As such, they will be tightly related to the irreducible integrands encountered in eq.~\eqref{eq:elliptic_integrals}. 
\item The kernels are linearly independent in the sense that there is no linear combination that evaluates to a total derivative. 
\item Since elliptic polylogarithms should have at most logarithmic singularities (but no poles), each $\varphi_{n}$ can have at most simple poles. 
\end{enumerate}
The explicit form of the integration kernels is discussed in the remainder of this section, together with some of the main properties of the iterated integrals $\textrm{E}_3$.

For $n=0$, we define 
\beq\label{eq:phi0_def}
\varphi_0(c,x) = \frac{c_3}{y}=\frac{c_3}{\sqrt{P_3(x)}}\,.
\eeq
The right-hand side of eq.~\eqref{eq:phi0_def} is independent of $c$, and we will often set $c=0$ in $\varphi_0$. The integral of $\varphi_0$ is closely related to the incomplete elliptic integral of the first kind $\EF$ in eq.~\eqref{eq:FEPi}. $\varphi_0$ defines a rational function on $\cE$ that is free of poles. Indeed, $\varphi_0$ has no poles for any finite value of $x$. Letting\footnote{The square in the change of variables is required because infinity is a branch-point and the integrand has a square root branch cut ending at $x=\infty$.} $x=1/u^2$, we see that there is no pole at infinity,
\beq
\int dx\,\varphi_{0}(0,x) = -2c_3\int du\,(1 + \ord(u))\,.
\eeq
The quantity $dx/y$ is often referred to as the \emph{holomorphic differential} on $\cE$.

While $\varphi_0$ is free of poles, the functions $\varphi_{\pm1}(c,x)$ have a simple pole at $x=c$,
\beq
\varphi_1(c,x) = \frac{1}{x-c} \textrm{~~and~~} \varphi_{-1}(c,x) = \frac{y_c}{y\,(x-c)}\,,
\eeq
where we define $y_c\equiv \sqrt{P_3(c)}$. Let us make some comments about these functions. First, using integration by parts we see that any integral involving $\varphi_{-1}$ with $c=a_i$ can be reduced to simpler integrals. Hence, the functions $\varphi_{-1}(a_i,x)$ are not part of our basis of integration kernels. Second, we see that the function $\varphi_1$ is independent of the branch points $a_i$, and it agrees with the integration kernel $f$ in eq.~\eqref{eq:MPL_kernel}. In other words, MPLs are a subset of the elliptic polylogarithms,
\beq\label{eq:E_to_G}
\Et{1 & \ldots & 1}{c_1 & \ldots& c_k}{x} = G(c_1,\ldots,c_k;x)\,.
\eeq
This is important in physics applications, where it often happens that the differential equations satisfied by an elliptic Feynman integral involve ordinary MPLs in the inhomogeneous term. 

Equation~\eqref{eq:elliptic_integrals} contains an additional integral which is not covered by the integration kernels defined so far, namely the integral over $x\,dx/y$. We have (using again $x = 1/u^2$)
\beq\label{eq:double_pole}
\int \frac{x\,dx}{y} = -\int du\,\left(\frac{2}{u^2}+\ord(u^0)\right)\,.
\eeq
The integrand in eq.~\eqref{eq:double_pole} has a double pole at infinity (or equivalently at $u=0$), so it violates our criterion of a basis of integration kernels with at most simple poles. We can, however, obtain an independent function with a simple pole by computing the primitive in eq.~\eqref{eq:double_pole}.
In addition, we have some freedom in how to define the basis element that corresponds to a double pole at infinity, because we can add any function without poles to the integrand, and in particular any multiple of the holomorphic differential. Here, we define
\beq\label{eq:Phi3_def}
\Phi_3(x,\vec a) = \widetilde{\Phi}_3(x,\vec a) -8c_3\,\frac{\eta_1}{\omega_1\,y} \,.
\eeq
The term proportional to the quasi-period $\eta_1$ is purely conventional at this point, and its role will only become clear in section~\ref{sec:torus_to_Weierstrass}. We define a primitive of $\Phi_3$ by
\begin{align}\label{eq:Z3_def}
Z_3(x,\vec a) &= \int_{a_3}^xdt\,\Phi_3(t,\vec a)\,.
\end{align}
Although the equation~\eqref{eq:cubic_eq} defining the elliptic curve is perfectly symmetric in the three branch points $a_i$, the choice of the lower integration boundary in eq.~\eqref{eq:Z3_def} breaks the permutation symmetry\footnote{The branch point at infinity would preserve the symmetry among the $a_i$. We prefer not to choose the point at infinity as an integration boundary, because $\Phi_3$ has a pole at infinity.}. A different choice of branch point as integration boundary only changes the value of the integral by a linear combination of the quasi periods, cf.~\eqref{eq:quasi-periods_def}.
The function $Z_3$ is regular everywhere, except at infinity where it has a simple pole,
\beq\label{eq:Z3_asymptotics}
Z_3(x) = \frac{2}{c_3}\sqrt{x} + \ord(1/\sqrt{x}) = \frac{2}{c_3\,u}+\ord(u^{-1})\,.
\eeq
We therefore include the function $Z_3$ into our basis of integration kernels with at most simple poles,
\beq
\varphi_{1}(\infty,x) = \frac{c_3}{y}\,Z_3(x)\,.
\eeq

To summarise, we see that for each $c \in {\mathbb{C}}\setminus\{a_1,a_2,a_3\}$ there are two basis elements $\varphi_{\pm1}(c,x)$ with a simple pole at $x=c$. If $c=a_i$, then only $\varphi_{1}(a_i,x)$ appears. One of the main differences between polylogarithmic functions on curves of genus zero and one is that, while in the case of the Riemann sphere it is possible to find a purely rational basis of integration kernels with at most simple poles, this is no longer true when working on an elliptic curve, and we need to include the transcendental function $Z_3$.

The fact that $Z_3$ is not rational has further consequences. In the case of ordinary MPLs, we can always reduce any product or power of the integration kernels $f$ in eq.~\eqref{eq:MPL_kernel} to a linear combination of integrals of the same kernels. For example, any product can be linearised using partial fractioning,
\beq\label{eq:partial_fractioning}
f(a,x)\,f(b,x) = \frac{1}{a-b}\,f(a,x) - \frac{1}{a-b}\,f(b,x)\,,
\eeq
and all higher powers can be recast in the form of a total derivative in $x$,
\beq
f(a,x)^k = \frac{1}{1-k}\,\partial_xf(a,x)^{k-1}\,,\quad k\neq 1\,.
\eeq
Similar relations, however, do not exist for the function $Z_3$, and so we need to complement our basis by all possible products and powers that involve this function.

Let us start by discussing the case of higher powers of $Z_3$. Since $Z_3$ has a simple pole at infinity, $Z_3^n$ has a pole of order $n$, and so it violates the condition that our integration kernels should have at most simple poles. We therefore need to include appropriate subtraction terms that allow us to remove the poles of higher order. While there is some arbitrariness in the precise definition of the subtraction terms, a suitable choice will be constructed in section~\ref{sec:torus_to_Weierstrass}. More precisely, based on the results of refs.~\cite{BrownLevin,Broedel:2014vla,MatthesThesis}, we define in section~\ref{sec:torus_to_Weierstrass} a family of polynomials $\cZ_n$ of degree $n$ such that $Z_3^{(n)}(x)\equiv \cZ_n(Z_3(x);x,y)$ is free of poles in $x$. The coefficients appearing in $\cZ_n$ are themselves polynomials in $(x,y)$. The explicit form of the polynomials $\cZ_n$ is rather involved, so we content ourselves to only present the explicit result for $Z_3^{(2)}$ and $Z_3^{(3)}$,
\beq\bsp\label{eq:Z2Z3_examples}
Z_3^{(2)}(x) &\,=\cZ_2\big(Z_3(x);x,y) =  \frac{1}{8}\,Z_3(x)^2-\frac{1}{2c_3^2}\,\left(x-\frac{s_1}{3}\right)\,,\\
Z^{(3)}_3(x) &\,=\cZ_3\big(Z_3(x);x,y) =  \frac{1}{96}\,Z_3(x)^3 - \frac{1}{8c_3^2}\,\left(x-\frac{s_1}{3}\right)\,Z_3(x) - \frac{1}{6}\,\frac{y}{c_3^3}\,.
\esp\eeq
Using the asymptotic behaviour of $Z_3$ in eq.~\eqref{eq:Z3_asymptotics}, it is easy to check that the previous expressions remain finite as $x\to\infty$ (Note that eq.~\eqref{eq:root3} implies $y<0$ for large $x$).
We define the integration kernels $\varphi_{n}(\infty,x)$, with $n\ge 2$, by
\beq\bsp\label{eq:higher_kernels}
\varphi_n(\infty,x)&\, = \frac{c_3}{y}\,Z_3^{(n)}(x)\,.
\esp\eeq
The previous equation remains valid for 
$n=1$ if we define $Z_3^{(1)}(x) = Z_3(x)$.
We expect that in applications to two-loop Feynman integrals only $Z_3^{(n)}$ for small values of $n$ show up, because the transcendental weight of a two-loop amplitude in four dimensions is constrained to be less or equal to four\footnote{The notion of weight used in physics is strictly speaking only defined for Feynman integrals that evaluate to ordinary MPLs. Since the weight of an MPL is closely connected to the number of integrations, we expect that the powers of $Z_3$ that can show up for small numbers of loops is rather limited.}.

Next, we need to consider products between $\varphi_{\pm1}(c,x)$ and powers of $Z_3$.
We define
\beq\bsp\label{eq:phi_n_def}
\varphi_n(c,x)&\, = \left(\frac{1}{x-c}+\frac{c_3}{2y}\,Z_3(x)\right)\,Z_3^{(n-1)}(x)\,,\\
\varphi_{-n}(c,x)&\,  = \frac{y_c}{y(x-c)}\,Z_3^{(n-1)}(x)\,.
\esp\eeq
Let us make some comments about these definitions.
Since $Z_3^{(n)}$ is free of poles for $n\ge2$, the integration kernels in eq.~\eqref{eq:phi_n_def} have at most simple poles. 
The term proportional to $Z_3(x)\,Z_3^{(n-1)}(x)$ is included to remove the pole at infinity of
$dx/({x-c})$. We use integration by parts to show that we do not need to consider kernels of the
form $\varphi_{-n}(a_i,x)$, because they can always be reduced to other classes of integrals.

To summarise, the integration kernels that define the elliptic polylogarithms in eq.~\eqref{eq:eMPLs_def} are given by the compact formulas (valid for $n>1$),
\beq\label{eq:final_Weierstrass}
\boxed{\bsp
\varphi_0(0,x)&\, = \frac{c_3}{y}\,,\\
\varphi_1(c,x)&\, = \frac{1}{x-c} \,,\qquad \varphi_{-1}(c,x) = \frac{y_c}{y(x-c)} \,,\qquad \varphi_1(\infty,x) = \frac{c_3}{y}\,Z_3(x)\,,\\
\varphi_n(c,x)&\, = \left(\frac{1}{x-c}+\frac{c_3}{2y}\,Z_3(x)\right)\,Z_3^{(n-1)}(x)\,,\\
\varphi_{-n}(c,x)&\,  = \frac{y_c}{y(x-c)}\,Z_3^{(n-1)}(x) \,,\qquad
\varphi_n(\infty,x)\, = \frac{c_3}{y}\,Z_3^{(n)}(x)\,.
\esp
}\eeq
The functions $\varphi_n(c,x)$ form the complete set of integration kernels needed to define
elliptic polylogarithms. For each $c\in{\mathbb{C}}\setminus\{a_1,a_2,a_3\}$ there are two infinite
families $\varphi_{\pm n}(c,x)$, $n\in \mathbb{N}$, with a simple pole at $x=c$. If $c=a_i$, $1\le
i\le4$ (with $a_4=\infty$), the family with negative index is reducible, and we only need to
consider the single infinite family $\varphi_{ n}(a_i,x)$. While it is manifest from
eq.~\eqref{eq:final_Weierstrass} that the $\varphi_{\pm n}$ have at most simple poles, we have not
yet shown that they satisfy the remaining two conditions spelled out at the beginning of this
section. In section~\ref{sec:torus_to_Weierstrass} we show that the integration kernels
$\varphi_{\pm n}$ are in one-to-one correspondence with the integration kernels that define the
elliptic polylogarithms considered in refs.~\cite{BrownLevin,Broedel:2014vla,MatthesThesis}. The latter are shown to define a complete and independent set of integration kernels in ref.~\cite{BrownLevin}, and so the same holds true for the functions defined in eq.~\eqref{eq:final_Weierstrass}.

Before proceeding further, we stress that some of the elliptic polylogarithms obtained by 
integrating once over the simplest 
kernels in eq~\eqref{eq:final_Weierstrass}, can be written in terms of the elliptic integrals in eq.~\eqref{eq:FEPi}.
We find, for $0<x<a_1<a_2<a_3<c$,
\beq\bsp\label{eq:E3_to_Pi}
\Et{0}{0}{x} &= -
\EF\! \left( \!\sqrt{\frac{x-a_1}{a_2-a_1}} \Big|  \lambda\! \right) - (x \leftrightarrow 0)\,,\\
\Et{-1}{c}{x} & = 
\frac{y_c}{c_3\,(c-{a_1})} \, \left[ 
 \Pi\!
   \left(\frac{{a_2}-{a_1}}{c-{a_1}}, \sqrt{\frac{x-a_1}{a_2-a_1}} \Big| \, \lambda\! \right)
- (x \leftrightarrow 0) \right]\,.
\esp\eeq
Similar relations can be obtained for other regions. Note that the signs in the previous formula are connected to our prescription for the branches of the square root in eq.~\eqref{eq:root3}.
We see from the previous formula that the elliptic integral of the third kind $\Pi$ plays a role very similar to to the logarithm in the case of the Riemann sphere.
We cannot give a closed analytical expression for the integral of $\varphi_1(\infty,x)$, because it requires integrating first
 over $Z_3(x)$, cf.~eq.~\eqref{eq:final_Weierstrass}. We can, however, provide an analytic
 expression for $Z_3(x)$ itself in terms of incomplete elliptic integrals. It is more convenient to do it in the region
$0<a_1<a_2<x<a_3$, where we have
\begin{align}\label{eq:Z3_to_E}
\!\!\!\!Z_3(x) &=  -4\, i\, \EE\!\left(\!\sqrt{\frac{x-a_3}{a_2-a_3}} \Big | 1-\lambda\! \right) + \frac{4i}{a_{13}} 
\left( a_1 - \frac{s_1}{3} + 8 c_3^2 \frac{\eta_1}{\omega_1}\right)
\EF\!\left(\!\sqrt{\frac{x-a_3}{a_2-a_3}}  \Big| 1-\lambda\! \right)\,.
\end{align}

Let us discuss some of the properties of the iterated integrals defined by eq.~\eqref{eq:final_Weierstrass}.
We have presented in detail the case of an elliptic curve defined by a cubic
polynomial, cf.~eq.~\eqref{eq:cubic_eq}. The results of this section are not
specific to cubic polynomials, and in section~\ref{sec:general} we define a
similar set of functions for elliptic curves defined by a quartic polynomial.
While the analytic form of some of the kernels, in particular of $\Phi_3$ and
$Z_3$, are different, the overall picture and the counting of the independent
integration kernels are the same: for each
$c\in\widehat{\mathbb{C}}\setminus\{a_1,\ldots,a_4\}$ there is a double
infinite family of integration kernels, while for each branch point there is a
single infinite family. 

The functions $\textrm{E}_3$ in eq.~\eqref{eq:eMPLs_def} satisfy all the basic properties of iterated integrals. In particular, they form a shuffle algebra
\beq
\textrm{E}_3(\vec c;x)\,\textrm{E}_3(\vec d;x) = \sum_{\vec w\in \vec c\shuffle\vec d}\textrm{E}_3(\vec w;x)\,,
\eeq
with $\vec c = \left(\begin{smallmatrix}n_1&\ldots& n_k\\ c_1&\ldots& c_k\end{smallmatrix}\right)$ and similarly for $\vec d$. At this point we have to make a comment about potential divergencies at $x=0$. Since the lower integration boundary in eq.~\eqref{eq:eMPLs_def} is $t=0$, the integral potentially diverges whenever $c_k=0$. We need to regularise this divergence (e.g., by introducing a suitable subtraction term), and there is a certain degree of arbitrariness in this procedure. However, we would like to do this in a way that preserves the shuffle algebra structure. In the case of ordinary MPLs, this is achieved by a modified integration contour, see eq.~\eqref{eq:log_def}. In appendix~\ref{app:shuffle_regularisation} we present a regularisation for $\textrm{E}_3$ that preserves the shuffle algebra structure.

Our paper is not the first to consider elliptic generalisations of MPLs~\cite{BrownLevin,Bloch:2013tra,Broedel:2014vla,MatthesThesis,Adams:2014vja,Adams:2015gva,Bloch:2014qca,Bloch:2016izu,Bonciani:2016qxi} or iterated integrals over integration kernels that involve square roots of polynomials of degree greater than two~\cite{Ablinger:2014bra}  have been considered. In particular, in refs.~\cite{BrownLevin,Broedel:2014vla,MatthesThesis} \emph{elliptic polylogarithms} have been defined as iterated integrals on a punctured torus, and a complete basis for the corresponding integration kernels on the torus has been defined. Moreover, it was proven in ref.~\cite{BrownLevin} that every iterated integral on the punctured torus can be expressed in terms of these functions. In the remainder of this paper we study the relationship between the elliptic polylogarithms of refs.~\cite{BrownLevin,Broedel:2014vla,MatthesThesis} and the iterated integrals defined in eq.~\eqref{eq:eMPLs_def}. One of our main results is that the iterated integrals $\textrm{E}_3$ are equivalent to the elliptic polylogarithms of refs.~\cite{BrownLevin,Broedel:2014vla,MatthesThesis}\footnote{Up to a technical distinction that will be discussed in more detail in section~\ref{sec:torus}.}, and we can cast every iterated integral $\textrm{E}_3$ as a linear combination of the elliptic polylogarithms of refs.~\cite{BrownLevin,Broedel:2014vla,MatthesThesis}, and vice-versa. A concrete algorithm how to perform this translation is presented in section~\ref{sec:torus_to_Weierstrass}.

Let us conclude this section by discussing another similarity between ordinary and elliptic polylogarithms, which is at the same time one of the main results of this paper.
At the end of section~\ref{sec:genus0} we have seen that the algebra $\cA_{\textrm{MPL}}$ generated by all MPLs with rational functions as coefficients is closed under both differentiation and integration. There is a similar result for the elliptic polylogarithms. Indeed, in ref.~\cite{BrownLevin} it was shown that every iterated integral on the punctured torus can be expressed in terms of elliptic poylogarithms (and rational functions). Since our iterated integrals define essentially the same class of functions as the elliptic polylogarithms of refs.~\cite{BrownLevin,Broedel:2014vla,MatthesThesis}, we conclude that one can evaluate all iterated integrals on an elliptic curve $\cE$ in terms of the iterated integrals $\textrm{E}_3$. 
More precisely, let $\cR_3\equiv \mathbb{C}(x,y)/\langle y^2=P_3(x)\rangle$ denote the field of rational functions of the elliptic curve $\cE$. We consider the algebra $\cA_3$ over $\cR_3$ generated by $Z_3(x)$ and all elliptic polylogarithms $\Et{n_1&\ldots n_k}{c_1&\ldots c_k}{x}$. Seen as a vector space over $\cR_3$, the algebra $\cA_3$ admits the presentation
\beq\label{eq:A3_def}
\cA_3 = \Big\langle Z_3^{(m)}(x)\,\Et{n_1&\ldots n_k}{c_1&\ldots c_k}{x}:m\ge 0, n_i\in\mathbb{Z}, c_i\in\widehat{\mathbb{C}}\Big\rangle_{\cR_3}\,.
\eeq
We call the quantity $l=m+k$ the \emph{total length}. The algebra $\cA_3$ shares many of the properties of $\cA_{\textrm{MPL}}$ of section~\ref{sec:genus0}. First, it is easy to check that $\cA_3$ is filtered by the total length\footnote{$\cA_3$ is not graded by the total length, nor by the weight.},
\beq
\cA_3 = \bigcup_{l=0}^\infty\cA_{3,l}\,,\textrm{  with  } \cA_{3,l_1}\cdot \cA_{3,l_2}\subseteq \cA_{3,l_1+l_2}\,.
\eeq
where elements of $\cA_{3,l}$ are linear combinations of the form
\beq\label{eq:linear_comb}
\sum_{\substack{m,\vec n,\vec c\\ m+|\vec n|\le l}}a_{m,\vec n,\vec c}(x,y)\,Z_3^{(m)}(x)\,\Et{\vec n}{\vec c}{x}\,.
\eeq
Second, $\cA_3$ is closed under differentiation with respect to $x$. If the coefficients in the linear combination~\eqref{eq:linear_comb} are constants, then differentiation lowers a non-zero total length by one unit, because it lowers the length of an elliptic polylogarithm, and the derivative of $Z_3$ is a rational function. In particular, we see that if $Z_3^{(m)}(x)\,\Et{\vec n}{\vec c}{x}$ has total length $m+|\vec n|$, then $\partial_x\Big(Z_3^{(m)}(x)\,\Et{\vec n}{\vec c}{x}\Big)$ has total length $m+|\vec n|-1$.
Finally, while the closure under differentiation is immediate, it is less obvious to see that $\cA_3$ is also closed under integration. More precisely,  for every $f(x)\in\cA_3$ there is a primitive $F(x)\in\cA_3$ such that $\partial_xF(x)=f(x)$. We provide an explicit and constructive proof of this statement in section~\ref{sec:algorithm}, where we present an algorithm to explicitly compute the primitive. This algorithm is in fact a generalisation of the classical algorithm to evaluate integrals of rational functions on elliptic curves reviewed at the beginning of this section, and it extends this classical algorithm to include elliptic polylogarithms and the function $Z_3$. Said differently, the classical algorithm allows one to find a primitive in the space $\cA_{3,0}$ of functions of total length zero (which are just rational functions), while our extension generalises it to functions of arbitrary total length.

%% file: torus.tex

\section{Elliptic curves and iterated integrals on a torus}
\label{sec:torus}
The aim of this section is to provide the necessary mathematical background to understand how the iterated integrals $\textrm{E}_3$ defined in the previous section are connected to the elliptic polylogarithms that appear in the mathematics literature~\cite{BrownLevin,MatthesThesis}. The material in this section is not new and is in principle well known. We include it nonetheless because we feel that some of these topics are rarely discussed in the Feynman integral literature.

\subsection{Elliptic functions}
Consider two complex numbers $\omega_1$ and $\omega_2$ that are linearly independent over the real numbers. For concreteness, we will assume that $\omega_1$ is real and positive while $\omega_2$ is purely imaginary with a positive imaginary part. We can define a \emph{lattice} (see fig.~\ref{fig:lattice})
\beq
\Lambda = \mathbb{Z}\,\omega_1 + \mathbb{Z}\,\omega_2 = \{m\,\omega_1+n\,\omega_2: m,n\in\mathbb{Z}\} \subset \mathbb{C}\,.
\eeq
Such a lattice is a discrete additive subgroup of $\mathbb{C}$. The torus associated to the lattice $\Lambda$ is defined as the quotient $\mathbb{C}/\Lambda$ of $\mathbb{C}$ by the lattice. In other words, we identify two complex numbers whenever they differ by an element from the lattice. The torus is then obtained by identifying opposite sites of the \emph{fundamental parallelgram} $\{r\,\omega_1+s\,\omega_2: 0\le r,s<1\}$.

\FIGURE[!t]{
\includegraphics{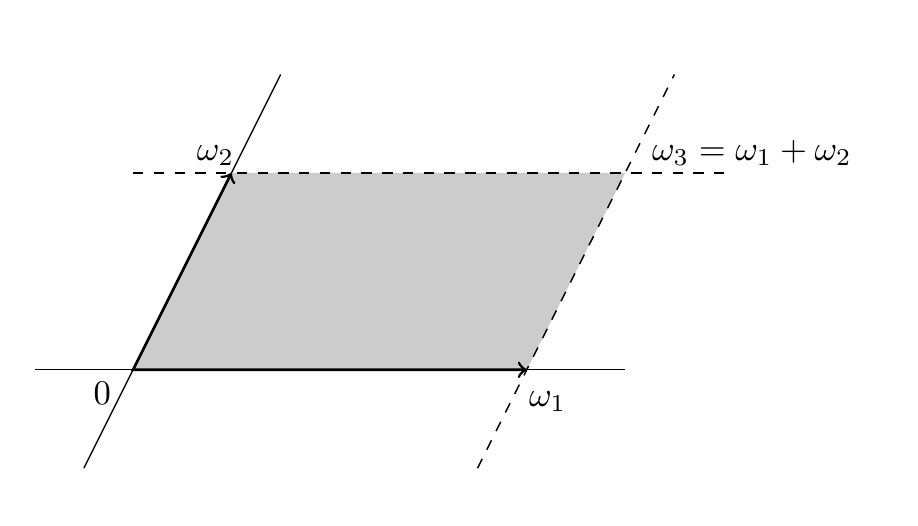}
\caption{\label{fig:lattice}The lattice $\Lambda$ spanned by the two periods $\omega_1$ and $\omega_2$. The grey-shaded area is the fundamental parallelogram.}
}

We now study functions on the torus $\mathbb{C}/\Lambda$. In order to be well-defined, any function on the torus must be invariant under translations by the \emph{periods} $\omega_i$, i.e., it must be a periodic function, $f(z+\omega_i)=f(z)$, $i=1,2$. An \emph{elliptic function} is a meromorphic periodic function. The singularity structure of an elliptic function is very constrained. In particular, every non-constant elliptic function must have at least two poles on the torus (counted with multiplicity, e.g., a double-pole counts as two poles), and the number of zeroes must equal the number of poles (again, counted with multiplicities). The prototypical example of an elliptic function is the \emph{Weierstrass $\wp$ function},
\beq\label{eq:wp_def}
\wp(z;\omega_1,\omega_2) = \frac{1}{z^2} + \sum_{(m,n)\neq(0,0)}\left(\frac{1}{(z+m\,\omega_1+n\,\omega_2)^2}-\frac{1}{(m\,\omega_1+n\,\omega_2)^2}\right)\,.
\eeq
We will always keep implicit the dependence of the Weierstrass $\wp$ function on the periods. The Weierstrass $\wp$ function is by construction periodic, and it has a double pole at every lattice point $z\in\Lambda$. In addition, it is an even function, $\wp(-z) = \wp(z)$.

The derivative of a periodic function is still periodic, and so the derivative of an elliptic function is itself elliptic. Since $\wp$ is even and has a double pole, its derivative must define an odd function with a triple pole (all definitions are understood modulo translations by the lattice). Hence, $\wp'$ must have three zeros, and these are located precisely at the \emph{half-periods} $\omega_i/2$, $i\in \{1,2,3\}$, with $\omega_3/2\equiv\omega_1/2+\omega_2/2$. Indeed, invariance under translations by $\Lambda$ gives
\beq
-\wp'(\omega_i/2) = \wp'(-\omega_i/2)=\wp'(\omega_i/2-\omega_i) = \wp'(\omega_i/2)\,,
\eeq
and so $\wp'(\omega_i/2) = 0$. 

The Weierstrass $\wp$ function and its derivative are not only the prototypical examples of functions that are both meromorphic and periodic, but they play a fundamental role in the theory of elliptic functions. In fact, they are sufficient to recover \emph{all} elliptic functions. More precisely, one can show that every elliptic function can be written as a rational function in $\wp$ and $\wp'$. Note that the set of all elliptic functions forms a field, and so we can identify the field of elliptic functions with the field of rational functions in $(\wp,\wp')$.

Finally, let us note that the Weierstrass $\wp$ function and its derivative are not independent, but they are coupled via a non-linear differential equation
\beq\label{eq:wp_deq}
\wp'^2 = 4\wp^3-g_2\,\wp-g_3 = 4(\wp-e_1)(\wp-e_2)(\wp-e_3)\,,
\eeq
where $g_i$ and $e_i$ are constants that depend on the two periods $\omega_1$ and $\omega_2$. 
By differentiation we see that all the higher derivatives of $\wp$ are polynomials in $(\wp,\wp')$, in agreement with the fact that every elliptic function is a rational function in $(\wp,\wp')$. For example, we have
\beq
\wp''(z) = \frac{1}{6}\,\wp(z)^2-\frac{1}{2}\,g_2\,.
\eeq

\subsection{From the torus to the elliptic curve: the Weierstrass model}
In section~\ref{sec:summary} we have considered elliptic curves given by a cubic equation of the form~\eqref{eq:cubic_eq}. In general, there can be several cubic polynomials that define the same elliptic curve $\cE$. In particular, it can be shown that via a judicious change of variables every elliptic curve can be represented as the solution set of a cubic equation of the form
\beq\label{eq:Weierstrass}
y^2 = 4x^3-g_2x-g_3 = 4(x-e_1)(x-e_2)(x-e_3)\,, \textrm{~~~~with~~~~} e_1+e_2+e_3=0\,.
\eeq
An equation of this form is called a \emph{Weierstrass equation} of the elliptic curve. We see that, 
upon identifying $(\wp,\wp')$ with $(x,y)$, eq.~\eqref{eq:wp_deq} has precisely the form of the Weierstrass equation, and so there is a strong connection between the Weierstrass $\wp$ function and elliptic curves.
The purpose of this section is to show that the torus and the Weierstrass $\wp$ function provide a natural parametrisation of any elliptic curve $\cE$.

Consider an elliptic curve $\cE$ with periods $\omega_1$ and $\omega_2$ given by the Weierstrass equation~\eqref{eq:Weierstrass}.
Comparing eq.~\eqref{eq:Weierstrass} to the differential equation~\eqref{eq:wp_deq} satisfied by the Weierstrass $\wp$ function, we see that for every point $z$ on the torus $\mathbb{C}/\Lambda$, the point $[\wp(z),\wp'(z),1]$ lies on the elliptic curve $\cE$. We may thus ask the converse question: given a point $[x,y,1]$ on $\cE$, can we find a point $z$ on the torus such that $[x,y,1] = [\wp(z),\wp'(z),1]$? It turns out that the answer to this question is positive. If $y=0$, we know that $\wp'$ vanishes on the half-periods, and so we have $[e_i,0,1] = [\wp(\omega_i/2),0,1]$. If $y\neq0$, then the function $f(z)\equiv \wp(z)-x$ is an elliptic function with a double pole at  $z=0$ on the torus. Since every elliptic function must have the same number of zeroes and poles, $f$ must have two zeroes (or a double zero) on the torus, and so the equation $\wp(z)=x$ has always two solutions, which differ by a sign because $\wp$ is even. Hence, there is a one-to-one mapping between the points of the torus $\mathbb{C}/\Lambda$ and the elliptic curve $\cE$, and the torus provides a canonical way to parametrise any elliptic curve. The map is explicitly given by 
\beq\label{eq:torus_to_curve}
z \mapsto [x,y,1] \equiv \left[\wp(z),\wp'(z),1\right]\,.
\eeq
Since $\wp$ has a pole at the origin, we see that the point $z=0$ on the torus is mapped to the point at infinity on the elliptic curve. The inverse map from the curve $\cE$ to the torus can also be given explicitly, but before we do so, we discuss in more detail what the map in eq.~\eqref{eq:torus_to_curve} implies for the structure of functions and integrals on the elliptic curve and the torus.

We start by analysing what elliptic functions correspond to under the map in eq.~\eqref{eq:torus_to_curve}.
We know that every elliptic function is a rational function in $(\wp,\wp')$. If $R$ is a rational function in two variables, then under the map in eq.~\eqref{eq:torus_to_curve} the elliptic function $R(\wp(z),\wp'(z))$ is mapped to the rational function $R(x,y)$ on the elliptic curve. In other words, under the map in eq.~\eqref{eq:torus_to_curve} the field of elliptic functions is mapped to the field of rational functions on $\cE$.

Next, we review some standard material on differential forms on an elliptic curve. An \emph{abelian differential} on $\cE$ is a differential one-form of the form $dx\,R(x,y)$, where $R$ is a rational function on $\cE$. It is customary to consider three different types of abelian differentials. An abelian differential of the first kind is holomorphic everywhere on $\cE$, and so in particular it has no poles on $\cE$. An abelian differential of the second kind is meromorphic, i.e., it is allowed to have poles on $\cE$, but the residue at every pole must vanish. Finally, a meromorphic differential with non-vanishing residues is called an abelian differential of the third kind.

Let us now discuss what happens to abelian differentials under the correspondence~\eqref{eq:torus_to_curve}. First, it follows from section~\ref{sec:plumber_review} that we can always reduce any abelian differential to a linear combination of the differentials in eq.~\eqref{eq:elliptic_integrals}. Let us analyse each of these differentials in turn. First, the differential $dx/y$ has no pole, and thus corresponds to a differential of the first kind. Under the map~\eqref{eq:torus_to_curve}, $dx/y$ corresponds to the standard holomorphic differential $dz$ on the torus. Indeed, letting $x=\wp(z)$ and using eq.~\eqref{eq:wp_deq}, we find 
\beq
\frac{dx}{y} = \frac{d\wp(z)}{\wp'(z)} = dz\,.
\eeq
Since the differential $x\,dx/y$ in eq.~\eqref{eq:double_pole} gives rise to a double pole without residue at infinity, it defines a differential of the second kind. On the torus it corresponds to the differential $\wp(z)\,dz = dz/z^2+\ord(z^0)$. Finally, the differential $dx/(y(x-c))$ has a simple pole at $x=c$ with residue $y_c\equiv\sqrt{P_3(c)}$, and so it defines a differential of the third kind. 

In general, it is easy to see that an abelian differential corresponds to a differential of the form $f(z)\,dz$ on the torus, where $f$ is an elliptic function. This in turn puts very strong restrictions on the pole structure of any abelian differential. Since an elliptic function must have at least two poles, it is not possible to find an abelian differential with only one simple pole. This is tightly connected to the fact that the primitive of an elliptic function is not periodic, because the choice of a lower integration boundary breaks the invariance under translations by a period. Instead, the primitive of an elliptic function defines a \emph{quasi-periodic} function, i.e., a function $F$ such that 
\beq\label{eq:quasi-periodic function}
F(z+\omega_i) = F(z) + C_i\,,
\eeq
where $C_i$ is a constant that may depend on the period $\omega_i$, but it is independent of $z$. If $f$ is an elliptic function, then any primitive of $f$ has the form
$F(z) = \int_{z_0}^zdz'\,f(z')$,
for some fixed point $z_0$. The function $F$ then satisfies eq.~\eqref{eq:quasi-periodic function} with $C_i = \int_{-\omega_i}^0dz'\,f(z')$.

The prototypical example of a quasi-periodic function is the \emph{Weierstrass zeta function}, defined as a primitive of the Weierstrass $\wp$ function (up to a sign). For $z$ inside the fundamental parallelogram it is given by the integral
\beq\label{eq:zeta_def}
\zeta(z) = \frac{1}{z}-\int_0^zdz'\,\left(\wp(z')-\frac{1}{z'^2}\right)\,.
\eeq
The Weierstrass zeta function is not invariant under lattice translations, but it transforms according to eq.~\eqref{eq:quasi-periodic function},
\beq
\zeta(z+\omega_i) = \zeta(z) + 2\eta_i\,,
\eeq
where $\eta_i\equiv\zeta(\omega_i/2)$ are the quasi-periods defined in eq.~\eqref{eq:quasi-periods_def}. We see from eq.~\eqref{eq:zeta_def} that the Weierstrass zeta function has a simple pole at $z=0$ (and hence at every point of the lattice $\Lambda$). As we will see, it is the primary building block to define differential one-forms with at most simple poles on the torus. As anticipated, this building block is no longer a periodic function, but it is only quasi-periodic.

Let us conclude this review with a discussion of the inverse of the map in eq.~\eqref{eq:torus_to_curve}. Consider a point $[a,b,1]$ on the elliptic curve, and for simplicity we assume $b>0$. Then we can find a point $z_{a}$ on the torus such that $(\wp(z_a),\wp'(z_a))=(a,b)$. The value of $z_a$ is given by \emph{Abel's map},
\beq\label{eq:Abel_map}
(a,b)\mapsto z_a\equiv c_3\int_{\infty}^{a}\frac{dx}{y}\mod\Lambda\,.
\eeq

\subsection{Elliptic polylogarithms}
In this section we review the construction of elliptic polylogarithms as iterated integrals of
differential one-forms with (at most) simple poles on the torus, following closely refs.~\cite{MatthesThesis,Broedel:2014vla,BrownLevin}. We define a class of iterated integrals by
\beq\label{eq:gamt_def}
\gamt{n_1 &\ldots& n_k}{z_1 & \ldots & z_k}{z} = \int_0^zdz\,g^{(n_1)}(z-z_1)\,\gamt{n_2 &\ldots& n_k}{z_2 & \ldots & z_k}{z}\,,
\eeq
where $z_i$ are complex numbers (assumed constant) and $n_i\in \mathbb{N}$ are positive integers. Note that in some cases the integral may require regularisation at $z=0$. In the following, we ignore this technical subtlety and refer to the literature on how to consistently define a regularised version of elliptic polylogarithms (cf., e.g., ref.~\cite{Broedel:2014vla}).

The integration kernels are defined through a generating series known as the \emph{Eisenstein-Kronecker series},
\beq\label{eq:Eisenstein-Kronecker}
F(z,\alpha) = \frac{1}{\alpha}\,\sum_{n\ge0}g^{(n)}(z)\,\alpha^n = \frac{1}{\alpha}\,\exp\left[-\sum_{j\ge 1}\frac{(-\alpha)^{j}}{j}\,(E_j(z)-G_j)\right]\,,
\eeq
where the quantities $E_j$ and $G_j$ in the exponential denote the Eisenstein functions and series respectively,\footnote{For $j=1,2$, these definitions require the Eisenstein summation convention,\begin{equation*}
\sum_{(m,n)\neq(0,0)}a_{mn}\equiv \lim_{N\to\infty}\lim_{M\to\infty} \sum_{n=-N}^N\sum_{m=-M}^ma_{mn}\,.
\end{equation*}}
\beq\label{eq:Eisenstein_def}
E_j(z) = \sum_{(m,n)\neq (0,0)}\frac{1}{(z+m\,\omega_1+n\,\omega_2)^j} {\rm~~and~~}G_j = \sum_{(m,n)\neq (0,0)}\frac{1}{(m\,\omega_1+n\,\omega_2)^j}\,.
\eeq
The Eisenstein series $G_j$ can be cast in the form of polynomials in the parameters $g_2$ and $g_3$ appearing in the Weierstrass equation.
From the series definition of $E_j$ we see that the Eisenstein functions satisfy the differential equation
\beq\label{eq:Eisenstein_deq}
\partial_zE_j(z)=-j\,E_{j+1}(z)\,.
\eeq
Moreover, comparing eq.~\eqref{eq:Eisenstein_def} to eq.~\eqref{eq:wp_def}, we see that $E_2(z) = \wp(z)+G_2$. Hence, for $j>2$ the Eisenstein functions can be expressed in terms of the derivatives of the Weierstrass $\wp$ function, 
\beq
E_j(z) = \frac{(-1)^j}{(j-1)!}\,\partial_z^{j-2}\wp(z)\,, \quad j>2\,.
\eeq
For $j=1$, instead, eq.~\eqref{eq:Eisenstein_deq} implies that $E_1$ is a primitive of the Weierstrass $\wp$ function, and so it is connected to the Weierstrass zeta function. More precisely, we have
\beq
E_1(z) = g^{(1)}(z) = \zeta(z) - \frac{2\eta_1}{\omega_1}\,z\,.
\eeq
We see that $g^{(1)}(z)$ has a simple pole at every lattice point, and in particular for $z=0$, but it is regular everywhere else. However, $g^{(1)}(z)$ is not invariant under translations by $\omega_2$, while it is for $\omega_1$, 
\beq\label{eq:g1_periodicity}
g^{(1)}(z+\omega_1) = g^{(1)}(z) \textrm{~~and~~} g^{(1)}(z+\omega_2) = g^{(1)}(z)-\frac{2\pi i}{\omega_1}\,.
\eeq

Since the Eisenstein functions for $j\ge2$ are elliptic functions, the exponential form of eq.~\eqref{eq:Eisenstein-Kronecker} implies that the functions $g^{(n)}$ for $n\ge 2$ can be expressed as polynomials of degree $n$ in $g^{(1)}$ whose coefficients are elliptic functions,
\beq\label{eq:cG_def}
g^{(n)}(z) = \cG_n\big(g^{(1)}(z);\wp(z),\wp'(z)\big)\,,
\eeq
where $\cG_n$ is a polynomial of degree $n$ in $g^{(1)}(z)$.  For example, we have
\beq\bsp\label{eq:g2g3_example}
g^{(2)}(z) &\,=  \cG_2\big(g^{(1)}(z);\wp(z),\wp'(z)\big) = \frac{1}{2}\,g^{(1)}(z)^2 - \frac{1}{2}\wp(z)\,,\\
g^{(3)}(z) &\,=  \cG_3\big(g^{(1)}(z);\wp(z),\wp'(z)\big) = \frac{1}{6}\,g^{(1)}(z)^3 - \frac{1}{2}\wp(z)\,g^{(1)}(z) - \frac{1}{6}\,\wp'(z)\,.
\esp\eeq
Note that the form of these polynomials is very reminiscent of the polynomials in eq.~\eqref{eq:Z2Z3_examples}. It follows from the exponential form of the Eisenstein-Kronecker series in eq.~\eqref{eq:Eisenstein-Kronecker} that the two leading coefficients in the polynomial $\cG_n$ are very simple constants,
\beq\label{eq:universal_coefficient}
\cG_n(g_1;\wp,\wp') = \frac{1}{n!}\,g_1^n + 0\times g_1^{n-1} + \ldots\,,
\eeq
where the dots indicate a polynomial in $g_1$ of degree $n-2$ at most. 

Let us discuss some properties of the functions $g^{(n)}$. First, it is easy to see that the $g^{(n)}$ are always invariant under translations by $\omega_1$, but not by $\omega_2$. Second, the $g^{(n)}$ are functions with definite parity,
\beq
g^{(n)}(-z) = (-1)^n\,g^{(n)}(z)\,,
\eeq
and satisfy the \emph{Fay identity}~\cite{BrownLevin},
\beq\bsp\label{eq:Fay}
g^{(m)}(z_1)\,g^{(n)}(z_2)&\, = -(-1)^n\,g^{(m+n)}(z_1-z_2)\\
&\, + \sum_{r=0}^n\binom{m+r-1}{m-1}\,g^{(n-r)}(z_2-z_1)\,g^{(m+r)}(z_1)\\
&\,+ \sum_{r=0}^m\binom{n+r-1}{n-1}\,g^{(m-r)}(z_1-z_2)\,g^{(n+r)}(z_2)\,.
\esp\eeq
The Fay identity is a generalisation of partial fractioning for the function $f$ appearing in the definition of ordinary MPLs, cf.~eq.~\eqref{eq:partial_fractioning}.
Finally, it can be shown that for $n\ge 2$, the $g^{(n)}$ are regular on the whole complex plane. For example, we have $g^{(1)}(z)=1/z + \ord(z^0)$ and $\wp(z)=1/z^2+\ord(z^0)$. Inserting this into eq.~\eqref{eq:g2g3_example}, we see that all the poles in $g^{(2)}(z)$ and $g^{(3)}(z)$ cancel.

Let us conclude this section by making a comment on the iterated integrals defined by
eq.~\eqref{eq:gamt_def} and the elliptic polylogarithms defined in
refs.~\cite{MatthesThesis,Broedel:2014vla,BrownLevin}. We have seen that the $g^{(n)}$ are not
periodic, and so they are strictly speaking not well-defined functions on the torus. This is
connected to the fact that there is no elliptic function with just a single simple pole, i.e., there
is no function that is both meromorphic and periodic with just one simple pole. Instead of giving up
periodicity to define the integration kernels $g^{(n)}$, we could also consider periodic functions
that are not meromorphic, i.e., that depend explicitly on the complex conjugate $\bar{z}$. This is
the approach taken in refs.~\cite{MatthesThesis,Broedel:2014vla,BrownLevin}, where elliptic polylogarithms are defined through the iterated integrals
\beq
\gam{n_1 &\ldots& n_k}{z_1 & \ldots & z_k}{z} = \int_0^zdz\,f^{(n_1)}(z-z_1)\,\gam{n_2 &\ldots& n_k}{z_2 & \ldots & z_k}{z}\,,
\eeq
where the functions $f^{(n)}$ are defined by the generating series
\beq\label{eq:Omega_def}
\Omega(z,\alpha) = \frac{1}{\alpha}\sum_{n\ge 0}f^{(n)}(z)\,\alpha^n = \exp\left[2\pi i\,\alpha\,\frac{\textrm{Im}\,z}{\omega_1\,\textrm{Im} \,\omega_2}\right]\,F(z,\alpha)\,.
\eeq
The functions $f^{(n)}$ have the same properties as the functions $g^{(n)}$, except that they are invariant under translations by both $\omega_1$ and $\omega_2$ and have an explicit dependence on the complex conjugate variable $\bar{z}$. Note that the dependence on the antiholomorphic variable is simple, because it only arises from the non-holomorphic exponential factor in eq.~\eqref{eq:Omega_def}.

In general, we have to give up either meromorphicity or periodicity in order to define elliptic polylogarithms. While in the mathematics and the string theory literature it is more common and natural to preserve periodicity, we prefer to work with functions that are manifestly meromorphic, at the expense of giving up periodicity. Our choice is motivated by the following considerations: integrands of Feynman integrals present themselves as purely meromorphic objects, so it feels unnatural to introduce an explicit dependence on the antiholomorphic variable. In addition, it is often much easier to work with meromorphic expressions. For example, it is not possible to integrate by parts in any naive way when working with non-meromorphic functions, because in that case $f(z,\bar{z})dz$ is not a total derivative, and so Stokes' theorem does not apply. Finally, we emphasise that in many practical applications the distinction between $\Gamma$ and $\widetilde{\Gamma}$ is immaterial. Indeed, whenever the integration contour is parallel to the real axis (which happens regularly in applications), the non-holomorphic exponential factor in eq.~\eqref{eq:Omega_def} is constant, and so $\Gamma$ and $\widetilde{\Gamma}$ are related in a trivial way. In addition, as we will see in the next section, in many cases of interest we can exchange $g^{(1)}$ with $f^{(1)}$ without loosing any information.

%% file: weierstrass_curve.tex

\section{Relating iterated integrals on the torus and on the elliptic curve}
\label{sec:torus_to_Weierstrass}
In the previous sections we have introduced two different classes of
generalisations of polylogarithms to elliptic curves: in
eq.~\eqref{eq:eMPLs_def} we have defined the functions $\textrm{E}_3$ as
iterated integrals on an elliptic curve with periods $\omega_1$ and $\omega_2$,
and in eq.~\eqref{eq:gamt_def} we have defined the functions
$\widetilde{\Gamma}$ as iterated integrals on the torus defined by the lattice
$\Lambda=\mathbb{Z}\,\omega_1+\mathbb{Z}\,\omega_2$. In this section we show
that these two sets of functions are just two different representations of the
same space of functions under the map in eq.~\eqref{eq:torus_to_curve} which
identifies the torus with the elliptic curve.

Before we can discuss how to relate the functions $\textrm{E}_3$ and
$\widetilde{\Gamma}$, we need to address the issue that the functions
$\textrm{E}_3$ have been defined for elliptic curves defined by arbitrary cubic
polynomials, cf.~eq.~\eqref{eq:cubic_eq}, while so far we have only defined a
map from the torus to an elliptic curve defined by a Weierstrass equation.
While we can always change coordinates and write any cubic equation in
Weierstrass form, it is convenient to define a map from the torus to
$\mathbb{CP}^2$ that lands us directly on the curve defined by the general
cubic equation in eq.~\eqref{eq:cubic_eq}. In the following we define such a
map, and we show that this map identifies the functions $\widetilde{\Gamma}$ on
the torus with the elliptic polylogarithms $\textrm{E}_3$.

Consider the elliptic function
\beq
\mu(z) = a_{31}\,\wp(z) + \frac{s_1}{3}\,.
\eeq
Using the properties of the Weierstrass $\wp$ function, one can check that $\mu$ satisfies the non-linear differential equation
\beq
\left(c_3\,\mu'\right)^2 = P_3(\mu) = (\mu-a_1)(\mu-a_2)(\mu-a_3)\,,
\eeq
where $c_3$ has been defined in eq.~\eqref{eq:c3_def}.
It is then easy to see that the map
\beq\label{eq:mu_map}
\mathbb{C}/\Lambda \to \cE;\qquad z\mapsto \left[\mu(z),c_3\,\mu'(z),1\right]\,,
\eeq
sends the torus to $\cE$. Using exactly the same argument as for the Weierstrass equation, we can show that this map is invertible. In particular, under this map the point $z=0$ is mapped to the point at infinity on $\cE$, while the half-periods $\omega_i/2$ map to the remaining branch points $a_i$,
\beq
\mu(\omega_1/2) = a_3\,,\qquad \mu(\omega_2/2) = a_1\,,\qquad  \mu(\omega_3/2) = a_2\,.
\eeq
The holomorphic differential $dx/y$ pulls back to the standard holomorphic
differential on the torus, $dx/y = dz/c_3$. Just like in the Weierstrass case,
every $c\neq a_i$ has two pre-images $\pm z_{c}$ on the torus such that
$\mu(\pm z_c)=c$. In the following, we assume without loss of generality that
$\textrm{Re}\,\mu'(z_c)>0$ (otherwise we exchange the roles of $+z_c$ and
$-z_c$). In the remainder of this section, we show that under the map in
eq.~\eqref{eq:mu_map} the functions $g^{(n)}$ and $\widetilde{\Gamma}$ of
refs.~\cite{MatthesThesis,Broedel:2014vla,BrownLevin} map to the functions
$Z_3^{(n)}$ and $\textrm{E}_3$ defined in section~\ref{sec:summary}.

Let $S$ be a finite set of points in $\widehat{\mathbb{C}}$. We consider
iterated integrals of rational functions that have poles at most at points in
$S$. For concreteness, we assume that $S$ contains $\infty$, and we define
$S'\equiv S\setminus\{\infty\}$, and it does not contain any of the zeroes
$a_i$ of $P_3$, nor the point $x=0$ (in order to avoid lengthy discussions
about the regularisation of eq.~\eqref{eq:eMPLs_def} -- see
appendix~\ref{app:shuffle_regularisation}). We stress that these assumptions
are not essential, but they allow us to avoid having to distinguish too many
different special cases. 
Our goal is to show that under the correspondence in
eq.~\eqref{eq:torus_to_curve} the complex vector space spanned by the
differential forms $dx\,\varphi_{n}(\infty,x)$ and $dx\,\varphi_{\pm n}(c,x)$,
$c\in S'$ (with $n\ge 0$), is identified with the vector space spanned by the
one-forms $dz\,g^{(n)}(z)$ and $dz\,g^{(n)}(z\pm z_c)$. A direct consequence of
this result is that the iterated integrals $\textrm{E}_3$ and
$\widetilde{\Gamma}$ define the same class of functions, and so the functions
$\textrm{E}_3$ coincide with the elliptic polylogarithms defined in the
mathematics literature (up to the different treatment of periodicity
vs.~memorphicity, cf.~section~\ref{sec:torus}).

Let us start by analysing the differential form $dx\,\varphi_{-1}(c,x)$. On the torus it corresponds to
\beq
dx\,\varphi_{-1}(c,x) = \frac{y_c\,dx}{y\,(x-c)} = dz\,\frac{\mu'(z_c)}{\mu(z)-\mu(z_c)} \equiv dz\, \alpha(z)\,.
\eeq
$\alpha$ is an elliptic function with simple poles only at $z=\pm z_c$ with residues $\pm1$. The difference
\beq\label{eq:alpha_diff}
\alpha(z) - \left[g^{(1)}(z-z_c) - g^{(1)}(z+z_c)\right]
\eeq
is then free of poles, and thus regular everywhere. Using eq.~\eqref{eq:g1_periodicity}, it is easy to check that the difference is periodic as a function of $z$, despite the fact that $g^{(1)}$ is not. Hence, the difference in eq.~\eqref{eq:alpha_diff} defines an elliptic function without any poles, and must therefore be constant. The constant is easily determined by using the fact that $\alpha(z)$ vanishes for $z=0$, and we find
\beq\label{eq:phi-1_to_g1}
dx\,\varphi_{-1}(c,x) = \frac{y_c\,dx}{y\,(x-c)} = dz\,\left[g^{(1)}(z-z_c) - g^{(1)}(z+ z_c) + 2g^{(1)}(z_c)\right]\,.
\eeq
We see that on the torus the differential form $dx\,\varphi_{-1}(c,x)$ corresponds to a linear combination with constant complex coefficients of the holomorphic differential $dz$ and the forms $dz\,g^{(1)}(z\pm z_c)$.

We can apply exactly the same reasoning to the differential form
\beq
dx\,\varphi_{1}(c,x) = \frac{dx}{x-c} = dz\,\frac{\mu'(z)}{\mu(z)-\mu(z_c)} \equiv dz\, \beta(z)\,.
\eeq
The only difference with respect to the previous case is that the residues at $z=\pm z_c$ are both $+1$, and there is also a simple pole at $z=0$ with residue $-2$, corresponding to the pole at infinity of $dx/(x-c)$.
We subtract the poles, and by exactly the same reasoning as before we conclude that the difference
\beq\label{eq:f1}
\beta(z) -\left[g^{(1)}(z-z_c) + g^{(1)}(z+z_c) -2 g^{(1)}(z)\right]
\eeq
must be constant. In order to determine this constant, we observe that both $\beta$ and eq.~\eqref{eq:f1} define an odd function, and so they must vanish at the origin. 
We then have
\beq\label{eq:phi1_to_g1}
dx\,\varphi_{1}(c,x) = \frac{dx}{x-c} = dz\,\left[g^{(1)}(z-z_c) + g^{(1)}(z+ z_c) - 2g^{(1)}(z)\right]\,.
\eeq

Let us now turn to $dx\,\varphi_1(\infty,x)$. If $x_0=\wp(z_0)$, eq.~\eqref{eq:Z3_def} gives
\beq\bsp\label{eq:Z1_to_g1}
Z_3(x_0)&\, = \int_{a_3}^{x_0}\frac{dx}{c_3\,y} \left(-x+\frac{s_1}{3}-8c_3^2\frac{\eta_1}{\omega_1}\right)\\
&\,=-4\int_{\omega_1/2}^{z_0}dz\,\left(\wp(z)+\frac{2\eta_1}{\omega_1}\right)\\
&\,=4\,g^{(1)}(z_0)\,.
\esp\eeq
Hence, we immediately find
\beq
dx\,\varphi_1(\infty,x) = \frac{c_3\,dx}{y}\,Z_3(x) = 4\,dz\,g^{(1)}(z)\,.
\eeq

To summarise, the one-forms $dx\,\varphi_{\pm 1}$ are linear combinations of the holomorphic differential and the one-forms $dz\,g^{(1)}$.
To complete the proof, we need to analyse what happens for $n>1$. In section~\ref{sec:summary} we have not given the complete definition of the polynomials $\cZ_{n}$ for $n>1$, but we have already noted the similarity for $n=2,3$ between the polynomials $\cZ_n$ in eq.~\eqref{eq:Z2Z3_examples} and $\cG_n$ in eq.~\eqref{eq:g2g3_example}. 
In general, we define
\beq\label{eq:Zn_to_gn}
Z_3^{(n)}(Z_3;x,y) \equiv {4}\,\cG_n\left(\frac{1}{4}\,Z_3;\frac{1}{4c_3^2}\Big(x-\frac{s_1}{3}\Big),\frac{y}{4c_3^3}\right)\,, \qquad n\ge 1\,.
\eeq
For $n=1$ we recover eq.~\eqref{eq:Z1_to_g1}. Equation~\eqref{eq:Zn_to_gn} immediately implies 
\beq
dx\,\varphi_n(\infty,x) = \frac{c_3\,dx}{y}\,Z_3^{(n)}(x) = 4\,dz\,g^{(n)}(z)\,.
\eeq
Next, let us turn to the one forms $dx\,\varphi_{-n}(c,x)$. Using eqns.~\eqref{eq:phi-1_to_g1} and~\eqref{eq:Zn_to_gn}, and applying the Fay identity~\eqref{eq:Fay}, we find
\beq\bsp\label{eq:phi-n_to_gn}
dx\,\varphi_{-n}(c,x) &\,= \frac{y_c\,dx}{y(x-c)}\,Z_3^{(n-1)}(x)\\
&\, = 4\,dz\,\left[g^{(1)}(z-z_c)-g^{(1)}(z+z_c)+ 2g^{(1)}(z_c)\right]\,g^{(n-1)}(z)\\
&\, = 4\,dz\,\Bigg[g^{(n)}(z-z_c)-g^{(n)}(z+z_c)+(1-(-1)^n)\,g^{(n)}(z_c)\\
&\,\phantom{=\,dz}\,+\sum_{r=1}^{n-1}g^{(n-r)}(z_c)\,\left(g^{(r)}(z-z_c)-(-1)^{n-r}g^{(r)}(z+z_c)\right)
\Bigg]\,.
\esp\eeq
We can identify $dx\,\varphi_{-n}(c,x)$ with the combination $dz\,\left[g^{(n)}(z-z_c)-g^{(n)}(z+z_c)\right]$, up to terms that involve less than $n$ powers of $g^{(1)}$. Applying exactly the same reasoning, we see that the corresponding combination with a plus sign is provided by $dx\,\varphi_{n}(c,x)$,
\begin{align}\label{eq:phi+n_to_gn}
dx\,\varphi_{n}(c,x) &\,= \left(\frac{dx}{x-c}+\frac{c_3\,dx}{2y}\,Z_3(x)\right)\,Z_3^{(n-1)}(x)\\
\nonumber&\, = 4\,dz\,\left[g^{(1)}(z-z_c)+g^{(1)}(z+z_c)\right]\,g^{(n-1)}(z)\\
\nonumber&\, = 4\,dz\,\Bigg[g^{(n)}(z-z_c)+g^{(n)}(z+z_c)+(1+(-1)^n)\,g^{(n)}(z_c)+2\,(n-1)g^{(n)}(z)\\
\nonumber&\,\phantom{=\,dz}\,+\sum_{r=1}^{n-1}g^{(n-r)}(z_c)\,\left(g^{(r)}(z-z_c)+(-1)^{n-r}g^{(r)}(z+z_c)\right)
\Bigg]\,.
\end{align}

To summarise, we have shown that every differential form $dx\,\varphi_{\pm n}$ can be written as a linear combination of the forms $dz\,g^{(k)}$ with complex coefficients that are constants with respect to $z$. As a consequence every iterated integral $\textrm{E}_3$ can be written as a linear combination with constant complex coefficients of elliptic polylogarithms $\widetilde{\Gamma}$, and vice-versa. The functions $\textrm{E}_3$ are simply an alternative linear basis for the shuffle algebra of elliptic polylogarithms. The change of basis is encoded into the relations~\eqref{eq:phi-1_to_g1}, \eqref{eq:phi1_to_g1}, \eqref{eq:Zn_to_gn} and~\eqref{eq:phi-n_to_gn}. For example, if $c$ is not a branch point, eq.~\eqref{eq:phi-1_to_g1} implies
\beq\label{eq:E-1_example}
\Et{-1}{c}{x} = \gamt{1}{z_c}{z} - \gamt{1}{-z_c}{z} + 2\,g^{(1)}(z_c)\,\gamt{0}{0}{z} - (z\leftrightarrow z_0)\,,
\eeq
where $z_0$ is the point on the torus such that $\wp(z_0)=0$ and $\wp'(z_0)>0$. Conversely, every elliptic polylogarithm $\widetilde{\Gamma}$ can be written as a linear combination of $\textrm{E}_3$ functions. For example, we have
\beq
\gamt{1}{z_c}{z} = \frac{1}{2}\,G(c;x) + \frac{1}{2}\,\Et{-1}{c}{x} - \frac{1}{4}\,Z_3(c)\,\Et{0}{0}{x} + \frac{1}{4}\,\Et{1}{\infty}{x}- (z\leftrightarrow \infty)\,.
\eeq

To conclude, let us make a comment about the connection between the iterated integrals $\widetilde{\Gamma}$ and $\Gamma$. The difference between the two sets of iterated integrals only lies in the fact that we have to give up either periodicity or holomorphicity to define the integration kernels (cf.~the discussion at the end of section~\ref{sec:torus}). In this section we have shown that we can write the iterated integrals $\textrm{E}_3$ in a natural way as linear combinations of elliptic polylogarithms $\widetilde{\Gamma}$. In some cases, however, we can also write them in terms of their periodic and non-holomorphic analogues $\Gamma$. Indeed, using eq.~\eqref{eq:g1_periodicity} it is easy to check that the combination of $g^{(1)}$ functions in eq.~\eqref{eq:phi-1_to_g1} is periodic with respect to both $\omega_1$ and $\omega_2$, and hence we can express it in terms of the periodic functions $f^{(1)}$,
\beq
dx\,\varphi_{-1}(c,x) = \frac{y_c\,dx}{y\,(x-c)} = dz\,\left[f^{(1)}(z-z_c) - f^{(1)}(z+ z_c) + 2f^{(1)}(z_c)\right]\,.
\eeq
The non-holomorphic contribution in $f^{(1)}$ cancels in the combination in the right-hand side. Similarly, we have
\beq
dx\,\varphi_{1}(c,x) = \frac{dx}{x-c} = dz\,\left[f^{(1)}(z-z_c) + f^{(1)}(z+ z_c) - 2f^{(1)}(z)\right]\,.
\eeq
In general, we see that we can write $\Et{n_1&\ldots&n_k}{c_1&\ldots&c_k}{x}$ as a linear combination of the periodic elliptic polylogarithms $\Gamma$ whenever $|n_i|\le 1$ and $c_i\neq \infty$. In particular, we can write the function $\Et{-1}{c}{x}$ in eq.~\eqref{eq:E-1_example} as
\beq\label{eq:E3-1_to_Gamma}
\Et{-1}{c}{x} = \gam{1}{z_c}{z} - \gam{1}{-z_c}{z} + 2\,f^{(1)}(z_c)\,\gam{0}{0}{z} - (z\leftrightarrow z_0)\,.
\eeq

%% file: algorithm.tex

\section{Iterated integrals on elliptic curves: an algorithmic approach}
\label{sec:algorithm}

In the previous section we have shown that the functions $\textrm{E}_3$ are in fact an alternative basis for the space of elliptic polylogarithms $\widetilde{\Gamma}$. In this section we consider integrals over elliptic polylogarithms multiplied by rational functions, and we present an algorithm to perform such integrals in terms of a well-defined class of functions. At the same time we prove the claim from section~\ref{sec:summary} that the algebra $\cA_3$ in eq.~\eqref{eq:A3_def} is closed under integration. 

Since $\cA_3$ is filtered by the total length, we can restrict the discussion to integrands with a given total length $l$. Using partial fractioning, we can reduce the problem to the following four types of integrals,
\beq\bsp\label{eq:ABCD_classes}
A_k[\cX] = \int dx\,x^k\,\cX(x)\,,\quad &B_{c,k}[\cX] = \int \frac{dx}{(x-c)^k}\,\cX(x)\,,\quad \\
C_k[\cX] = \int \frac{dx}{y}\,x^k\,\cX(x)\,,\quad &D_{c,k}[\cX] = \int \frac{dx}{y\,(x-c)^k}\,\cX(x)\,,
\esp\eeq
where $\cX(x) = Z_3^{(m)}(x)\,\Et{\vec n}{\vec c}{x}$ has total length $m+|\vec n|=l$.
If $l=0$, we recover the classical case of integrating a rational function on an elliptic curve (see section~\ref{sec:plumber_review}).
Not all integrals in a given family are independent, but they are related by integration by parts. We discuss each family in turn. 

The integrals in the $A$-family satisfy the recursion\footnote{We do not include boundary terms in the IBPs, because we are only interested in primitives.},
\beq
\frac{x^{k+1}}{k+1}\,\cX(x) = \frac{1}{k+1}\,A_{k+1}[\partial_x\cX] + A_{k}[\cX]\,.
\eeq
Note that the total length of $\partial_x\cX$ is $l-1$.
The recursion therefore allows us to increase the value of $k$ while at the same time to lower the value of the total length. We can therefore recursively lower the value of $l$, until we reach $l=0$ and the integral is elementary. However, the recursion has a singularity for $k=-1$, and so we cannot reduce integrals of the type $A_{-1}[\cX]$.

For integrals of type $B$, we only need to consider the case $k>0$. We have the recursion
\beq
\frac{\cX(x)}{(k-1)\,(x-c)^{k-1}} = \frac{1}{k-1}\,B_{c,k-1}[\partial_x\cX] + B_{c,k}[\cX]\,.
\eeq
We can lower the value of $k$, and at the same time lower the total length. The recursion has a singularity for $k=1$, and so we are left with integrals of the form $B_{c,1}[\cX]$ that cannot be reduced.

Integrals of type $C$ satisfy the recursion
\beq\bsp
y\,x^{k-2}\,\cX(x) &\,=\frac{1}{2}\sum_{l=0}^3(-1)^l\,s_l(a_1,a_2,a_3)\,\left[(2k-1-l)\,C_{k-l}[\cX]
+2\,C_{k+1-l}[\partial_x\cX]\right]\,.
\esp\eeq
The recursion is non-singular for all integer values of $k$. It has depth three, and so we can express all integrals from the family $C$ through three representatives of this family. We choose these irreducible integrals to correspond to $C_k[\cX]$, $k\in\{-1,0,1\}$.

Finally, the integrals in family $D$ satisfy the recursion
\begin{align}\label{eq:D3_rec}
&\frac{y}{(x-c)^{k-1}}\,\cX(x) = 
-\frac{1}{2}\sum_{l=0}^3\left[(2k-2-l)\,D_{c,k-l}[\cX]-2\,D_{c,k-1-l}[\partial_x\cX]\right]\,\frac{1}{l!}\partial_c^ly_c^2
 \,.
\end{align}
The recursion has depth three. However, for $k\le0$, the integrals reduce to integrals of type $C$, and so we choose as many of our basis integrals as possible to lie in the family $C$. The only obstacle to this is the appearance of a singularity for $k=1$ in the recursion. Hence, every integral in the family $D$ can be reduced to the integrals $D_{c,1}[\cX]$, as well as integrals of type $C$.

The previous discussion only applies if $c$ is not a zero of $P_3$. Indeed, if for example $c=a_1$, the coefficient of $D_{c,k}$ in eq.~\eqref{eq:D3_rec} vanishes, and so we cannot use the recursion to relate $D_{c,k}$ to other integrals. Instead, we have
\begin{align}
&\frac{y}{(x-a_1)^{k}}\,\cX(x)  =-\frac{1}{2}\sum_{l=0}^2\left[(2k-1-l)\,D_{a_1,k-l}[\cX]-2\,D_{a_1,k-1-l}[\partial_x\cX]\right]\,\frac{1}{l!}\partial_{a_1}^l(a_{12}a_{13})
\,.
    \end{align}
In this case the recursion is non-singular for all integer values of $k$, and so all integrals of the type $D_{a_1,k}[\cX]$ can be reduced to integrals of type $C$, in agreement with the fact that we do not need to consider integration kernels $\varphi_{-1}(a_1,x)$.

To summarise, using integration by parts we can reduce all integrals to the following six classes of integrals,
\begin{align}\label{eq:ABCD_MIs}
A_{-1}[\cX] &\,= \int \frac{dx}{x}\,\cX(x)\,,\qquad B_{c,1}[\cX] = \int \frac{dx}{x-c}\,\cX(x)\,,\qquad C_{1}[\cX] = \int \frac{x\,dx}{y}\,\cX(x)\,,\\
\nonumber
C_{0}[\cX] &\,= \int \frac{dx}{y}\,\cX(x)\,,\qquad C_{-1}[\cX] = \int \frac{dx}{y\,x}\,\cX(x)\,,\qquad D_{c,1}[\cX] = \int \frac{dx}{y\,(x-c)}\,\cX(x)\,.
\end{align}
It is easy to see that these integrals are in one-to-one correspondence with the differential forms $\varphi_{\pm n}$, and they can be performed using the definition of the iterated integrals $\textrm{E}_3$ in eq.~\eqref{eq:eMPLs_def}. The only case that needs some explanation are the integrals $C_1[Z^{(m)}_3(x)\,\textrm{E}_3]$. Equivalently, we may consider the integrals
\beq
\widetilde{C}_1[Z^{(m)}_3\,\textrm{E}_3] = \int dx\,\Phi_3\,Z^{(m)}_3\,\textrm{E}_3\,.
\eeq
In the previous equation, and until the end of this section, we keep the dependence of all quantities on $x$ implicit.
Since $\Phi_3$ has a double pole at infinity, it is not part of our basis of integration kernels, but its primitive $Z_3$ is. We can integrate by parts and we obtain
\beq
\widetilde{C}_1[Z^{(m)}_3\,\textrm{E}_3] = Z_3\,Z^{(m)}_3\,\textrm{E}_3 - \int dx\,Z_3\,\left(\textrm{E}_3\,\partial_xZ^{(m)}_3+ Z^{(m)}_3\,\partial_x\textrm{E}_3\right)\,.
\eeq
We have, with $m>1$,
\beq
Z_3\,\textrm{E}_3\,\partial_xZ^{(m)}_3 = \frac{4^{1-m}}{(m-1)!}\,Z_3^m\,\Phi_3\,\textrm{E}_3+\ldots = m\,\Phi_3\,Z^{(m)}_3\,\textrm{E}_3+\ldots\,,
\eeq
and so we reproduce the original integral that we started from. Hence, the total length has not been lowered, and so we cannot apply our recursive argument based on the total length. In the following we show how this integral can be evaluated. We only discuss the case where $\textrm{E}_3$ is absent from the integrand, because the argument is independent of the appearance of $\textrm{E}_3$. Using eq.~\eqref{eq:universal_coefficient}, we can write
\beq
Z_3^{(m)} =\frac{4^{1-m}}{m!}\,Z_3^m + Z_R^{(m)}\,,\qquad m>1\,,
\eeq
where $Z_R^{(m)}$ is a polynomial of degree at most $m-1$ in $Z_3$,
and we find
\beq\bsp
\int dx\,\Phi_3\,Z_3^{(m)} &\,= Z_3\,Z_3^{(m)} - \int dx\, Z_3\,\partial_xZ_3^{(m)}\\
&\,= Z_3\,Z_3^{(m)} - \int dx\, Z_3\,\left[\frac{4^{1-m}}{(m-1)!}\,Z_3^{m-1}\,\Phi_3+\partial_xZ_R^{(m)}\right]\\
&\,= Z_3\,Z_3^{(m)} - \int dx\, Z_3\,\partial_xZ_R^{(m)}-m\,\int dx\,\Phi_3\,\left[Z_3^{(m)}-Z_R^{(m)}\right]\,.
\esp\eeq
Hence, we have
\beq\bsp
(1+m)\,\int dx\,\Phi_3\,Z_3^{(m)}  = Z_3\,Z_3^{(m)} - \int dx\, Z_3\,\partial_xZ_R^{(m)}+m\,\int dx\,\Phi_3\,Z_R^{(m)}\,.
\esp\eeq
The right-hand side only involves polynomials in $Z_3$ of degree strictly less than $m$, i.e., it has a lower total length, and so recursively we know how to do these integrals.

%% file: general.tex

\section{Elliptic curves defined by quartic polynomials}
\label{sec:general}

So far we have only discussed elliptic curves that are defined through a cubic equation, cf.~eq.~\eqref{eq:cubic_eq}, and we have not yet discussed what happens in the case of elliptic curves defined by a quartic polynomial. The quartic case is particularly relevant for physics, because, e.g., the maximal cut of the sunrise integral leads to an elliptic curve defined by a quartic polynomial (cf., e.g., refs.~\cite{Broadhurst:1987ei,Bauberger:1994by,Bauberger:1994hx,Laporta:2004rb}). Since every elliptic curve can be represented as the zero set of some cubic equation (e.g., its Weierstrass equation) via a suitable change of variables, the results of the previous sections are in principle sufficient to cover all possible elliptic curves. In practise, however, the change of variables to the cubic form can be rather cumbersome, and it may be preferable to have a formulation where one can directly work with general quartic polynomials. In this section we show that this can be achieved, and we formulate all the results of the previous sections for elliptic curves defined by general quartic polynomials,
\beq\label{eq:y2=P4}
y^2 = P_4(x) = (x-a_1)(x-a_2)(x-a_3)(x-a_4)\,.
\eeq
We assume again that the roots $a_i$ are real and ordered according to $a_1<a_2<a_3<a_4$, and we choose the branches of the square root as follows,
\beq\bsp
\sqrt{P_4(x)}&\,\equiv\sqrt{|P_4(x)|}\,\big[-\theta(x\le a_1)-i\,\theta(a_1<x\le a_2)+ \theta(a_2<x\le a_3) \\
&\,\phantom{\equiv\sqrt{|P_4(x)|}\,[}+i\,\theta(a_3<x\le a_4) -\theta(a_4<x)\big]\\
&\,=\sqrt{|P_4(x)|}\times\left\{\begin{array}{ll}
-1\,,& x\le a_1\textrm{ or }x > a_4\,,\\
-i\,,& a_1<x\le a_2\,,\\
\phantom{-}1\,,& a_2<x\le a_3\,,\\
\phantom{-}i\,,& a_3<x\le a_4\,.
\end{array}\right.
\esp \label{eq:rsigns}
\eeq
With this convention the periods can be written as
\beq
\omega_1 = 2\,c_{4}\int_{a_2}^{a_3}\frac{dx}{y} = 2\,\EK(\lambda) 
\textrm{~~~~~~and~~~~~~}\omega_2 = 2\,c_{4}\int_{a_1}^{a_2}\frac{dx}{y} = 2i\,\EK(1-\lambda)\,,
\label{eq:periods4}
\eeq
with 
\beq\label{eq:lambda4}
\lambda = \cra(a_1,a_4,a_3,a_2) = \frac{a_{14}\,a_{23}}{a_{13}\,a_{24}} \qquad \textrm{and}\qquad
c_{4} = \frac{1}{2}\sqrt{a_{13}a_{24}}\,,
\eeq
and the $j$-invariant is given by eq.~\eqref{eq:j-invariant}. We stress here that eq.~\eqref{eq:rsigns}
implies a negative sign for the square-root in the region $a_2<x\leq a_3$, which in turn 
determines the overall sign of $\omega_2$ in eq.~\eqref{eq:periods4}.

Let us now discuss the abelian differentials of the second kind on $\cE$. The differential $x\,dx/y$, which provided the differential of the second kind in the cubic case (cf.~eq.~\eqref{eq:Phi_3_tilde}), is no longer a good candidate. Indeed, letting $u=1/x$, we see that 
\beq\label{eq:quartic_simple_pole_infty}
\int\frac{x\,dx}{y} = \int{du}\,\left[-\frac{1}{u}+\ord(u^0)\right]\,.
\eeq
Equation~\eqref{eq:quartic_simple_pole_infty} reveals that $x\,dx/y$ has a simple pole at infinity, and hence it defines a differential of the third kind. Instead, a valid differential of the second kind in the quartic case is\footnote{We will show in section~\ref{sec:algorithm_4} that, unlike in the cubic case, $x^2\,dx/y$ cannot be reduced using integration-by-parts identities, and so it defines a genuine master integral in the quartic case.}
\beq\label{eq:quartic_2nd_kind}
\frac{x^2\,dx}{y} - \frac{s_1}{2}\,\frac{x\,dx}{y}\,.
\eeq
We emphasise that it is mandatory to consider this particular linear combination. Indeed, $x^2\,dx/y$ is by itself not a valid differential of the second kind, because it has non-vanishing residue at infinity,
\beq
\int\frac{x^2\,dx}{y} = \int du\,\left[-\frac{1}{u^2}-\frac{s_1}{2\,u}+\ord(u^0)\right]\,.
\eeq
We can add any multiple of the holomorphic differential to eq.~\eqref{eq:quartic_2nd_kind}, and we still obtain a differential of the second kind. We find it convenient to define
\beq\label{eq:tilde_Phi_4_def}
\widetilde{\Phi}_4(x,\vec a) \equiv \frac{1}{c_4\,y} \left( x^2 - \frac{s_1}{2}\,x + \frac{s_2}{6} \right)\,.
\eeq
As usual, we will keep the dependence on $\vec a$ implicit.
The quasi-periods can then be expressed as elliptic integrals of the differential of the second kind (cf.~eq.~\eqref{eq:quasi-periods_def}),
\begin{align}
\eta_1 &\,= -\frac{1}{2}\int_{a_2}^{a_3}dx\,\widetilde{\Phi}_4(x) = \EE(\lambda) -\frac{2-\lambda}{3}\EK(\lambda)\,,\\
\nonumber\eta_2 &\,= -\frac{1}{2}\int_{a_1}^{a_2}dx\,\widetilde{\Phi}_4(x)= -i\,\EE(1-\lambda) +i\,\frac{1+\lambda}{3}\,\EK(1-\lambda)\,.
\end{align}

\subsection{From the torus to the elliptic curve}
Just like in the case of an elliptic curve defined by a quartic polynomial, we can define a map from the torus to the elliptic curve $\cE$,
\beq\label{eq:kappa_map}
\mathbb{C}/\Lambda \to \cE;\qquad z\mapsto \left[\kappa(z),c_{4}\kappa'(z),1\right]\,,
\eeq 
where $\kappa$ denotes the elliptic function
\beq\label{eq:kappa_def}
\kappa(z) = \frac{-3a_1a_{13}a_{24}\wp(z) + a_1^2\bar{s}_1-2a_1\bar{s}_2+3\bar{s}_3}{-3a_{13}a_{24}\wp(z) + 3a_1^2-2a_1\bar{s}_1+\bar{s}_2}\,,
\eeq
with $\bar{s}_n\equiv s_n(a_2,a_3,a_4)$. Using the differential equation~\eqref{eq:wp_deq} satisfied by the Weierstrass $\wp$ function, it is easy to check that $\kappa$ satisfies the non-linear differential equation
\beq
\left(c_{4}\,\kappa'\right)^2 = P_4(\kappa)=(\kappa-a_1)(\kappa-a_2)(\kappa-a_3)(\kappa-a_4)\,.
\eeq
Hence the image of the torus under the map in eq.~\eqref{eq:kappa_map} is precisely the elliptic curve defined by the quartic equation~\eqref{eq:y2=P4}. The holomorphic differential $dx/y$ on $\cE$ corresponds to $dz/c_{4}$ on the torus, and the inverse of eq.~\eqref{eq:kappa_map} is given by a variant of Abel's map in eq.~\eqref{eq:Abel_map}
\beq\label{eq:Abel_map_quartic}
(x_0,y_0)\mapsto z_0\equiv c_{4}\int_{a_1}^{x_0}\frac{dx}{y}\mod\Lambda\,.
\eeq

The half periods are naturally mapped by $\kappa$ to the zeroes of the polynomial $P_4$,
\beq\label{eq:kappa_half_periods}
\kappa(0) = a_1\,,\quad \kappa(\omega_1/2) = a_4\,,\quad \kappa(\omega_2/2) = a_2\,,\quad \kappa(\omega_3/2) = a_3\,.
\eeq
At this point we see a main difference between the cubic and quartic cases: while for a cubic polynomial the point at infinity is always a branch point, this is no longer the case for a quartic polynomial, and so the point at infinity of $\cE$ is not the image under $\kappa$ of a half-period. Indeed, we see from eq.~\eqref{eq:kappa_half_periods} that $\kappa$ is regular for every half-period, and it is singular only when the denominator in eq.~\eqref{eq:kappa_def} vanishes. Hence, there must be two points $\pm z_{\ast}$ on the torus such that denominator in eq.~\eqref{eq:kappa_def} vanishes. The value of $z_{\ast}$ is determined by Abel's map in eq.~\eqref{eq:Abel_map_quartic},
\beq
z_{\ast} \equiv c_{4}\int_{a_1}^{\infty}\frac{dx}{y}\mod\Lambda\,.
\eeq

The point $z_{\ast}$ plays an important role in understanding the structure of differentials on $\cE$ that have poles at infinity. We know for example that the differential of the second kind $dx\,\widetilde{\Phi}_4$ has a double pole at infinity, and so it must have double poles at $z=\pm z_{\ast}$ on the torus. In the following we derive a formula which makes this explicit. We obviously have
\beq
dx\,\widetilde{\Phi}_4(x) = \frac{dz}{c_{4}^2}\,\left[\kappa(z)^2 - \frac{s_1}{2}\,\kappa(z)+\frac{s_2}{6}\right]\,.
\eeq
We know that $\kappa(z)$ has poles at $z=\pm z_{\ast}$.
Using the explicit definition of $\kappa$ in eq.~\eqref{eq:kappa_def}, it is easy to show that
\beq
\kappa(z)^2 - \frac{s_1}{3}\,\kappa(z) = \frac{c_{4}^2}{(z\mp z_{\ast})^2} + \ord(z\mp z_{\ast})^0\,.
\eeq
Since the Weierstrass $\wp$ function has a double pole at the origin, we conclude that the elliptic function 
\beq
\alpha(z)\equiv \kappa(z)^2 - \frac{s_1}{2}\,\kappa(z) -c_{4}^2\,\left[\wp(z-z_{\ast}) + \wp(z+z_{\ast})\right]
\eeq
is free of poles and thus constant. The value of the constant is
\beq
\alpha(0) = a_1^2 - a_1\,\frac{s_1}{2} -2\,c_{4}^2\,\wp(z_{\ast}) = -\frac{s_2}{6}\,,
\eeq
where the value of $\wp(z_{\ast})$ is determined by the requirement that the denominator in eq.~\eqref{eq:kappa_def} vanishes.
We thus have
\beq\label{eq:Phi4_wp}
dx\,\widetilde{\Phi}_4(x) = dz\,\left[\wp(z-z_{\ast}) + \wp(z+z_{\ast})\right]\,.
\eeq
The previous formula makes explicit the fact that $dx\,\widetilde{\Phi}_4$ has a double pole at infinity on $\cE$, and thus double poles at $z=\pm z_{\ast}$ on the torus. We can apply exactly the same reasoning to the differential of the third kind with a simple pole at infinity, and we find
\beq\label{eq:kappa_g1}
\frac{x\,dx}{y} = \frac{dz}{c_{4}}\,\kappa(z)  = \frac{a_1\,dz}{c_{4}} + dz\,\left[g^{(1)}(z-z_{\ast}) - g^{(1)}(z+z_{\ast}) + 2g^{(1)}(z_{\ast})\right]\,.
\eeq
We see that the right hand side has simple poles both at $z=z_{\ast}$ and $z=-z_{\ast}$, in agreement with the fact that an elliptic function cannot have a single simple pole.

\subsection{Elliptic polylogarithms associated to curves defined by a quartic equation}
The elliptic polylogarithms attached to an elliptic curve defined by a quartic equation are defined in complete analogy to the cubic case in eq.~\eqref{eq:eMPLs_def},
\beq\label{eq:E4_def}
\Efe{n_1 & \ldots & n_k}{c_1 & \ldots& c_k}{x}{\vec{a}} = \int_0^xdt\,\psi_{n_1}(c_1,t,\vec a)\,\Efe{n_2 & \ldots & n_k}{c_2 & \ldots& c_k}{t}{\vec a}\,,
\eeq
with $n_i\in\mathbb{Z}$ and $c_i\in\widehat{\mathbb{C}}$, and the recursion starts with $\textrm{E}_4(;x,\vec a)=1$. The branch points are encoded in the vector $\vec a=(a_1,a_2,a_3,a_4)$. We will keep the dependence of all quantities on $\vec a$ implicit from now on. The integration kernels $\psi_n$ take a form very similar to the kernels that appear in the cubic case, cf.~eq.~\eqref{eq:final_Weierstrass}. We can write for $n \ge 2$ 
\beq\label{eq:final_quartic}
\boxed{\bsp
\psi_0(0,x) &\,= \frac{c_4}{y}\,,\\
\psi_1(c,x)&\, = \frac{1}{x-c}\,,\qquad \psi_{-1}(c,x) = \frac{y_c}{y(x-c)}\,, \\
\psi_1(\infty,x)&\, = \frac{c_4}{y}\,Z_4(x)\,,\qquad \psi_{-1}(\infty,x) = \frac{x}{y}\,, \\
\psi_{-n}(\infty,x)&\, = \frac{x}{y}\,Z_4^{(n-1)}(x)-\frac{\delta_{n2}}{c_4}\,,\\
\psi_n(c,x)&\, = \frac{1}{x-c}\,Z_4^{(n-1)}(x)-\delta_{n2}\,\Phi_4(x)\,,\\
\psi_n(\infty,x) &\,= \frac{c_4}{y}\,Z_4^{(n)}(x)\,,\qquad \psi_{-n}(c,x)  = \frac{y_c}{y(x-c)}\,Z_4^{(n-1)}(x)\,,
\esp
}\eeq
where we have defined  $Z_4^{(0)}(x)\equiv 1$, 
\beq
\Phi_4(x)\equiv\widetilde\Phi_4(x) +4c_{4}\, \frac{\eta_1}{\omega_1}\,\frac{1}{y}\,,
\eeq
and $Z_4(x) = Z_4^{(1)}(x)$ is a primitive of $\Phi_4(x)$,
\beq
Z_4(x) \equiv \int_{a_1}^xdx'\,\Phi_4(x')\,.
\eeq
Note that the choice of the lower integration boundary again breaks the symmetry among the branch points $a_i$.
It is easy to check that $Z_4$ has a simple pole at infinity, and it is regular everywhere else.
The functions $Z_4^{(n)}$ for $n>1$ are defined via the polynomials $\cG_n$ introduced in eq.~\eqref{eq:cG_def},
\beq\label{eq:Z4n_def} 
Z^{(n)}_4(x) \equiv \cG_n\!\left(\!-\frac{1}{2}\left(Z_4(x)-\frac{y}{c_4(x-a_1)}\right);p_1(x),p_2(x,y)\!\right) \,,
\eeq
with 
\beq\bsp
p_1(x) &\,= \frac{a_1^2\,\bar{s}_1-2a_1\,\bar{s}_2+3\bar{s}_3-(3a_1^2-2a_1\,\bar{s}_1+\bar{s}_2)\,x}{3\,a_{13}\,a_{24}\,(x-a_1)}\,,\\
p_2(x,y)&\,=-\frac{a_{12}\,a_{14}}{a_{24}\,c_{4}}\,\frac{y}{(x-a_1)^2}\,.
\esp\eeq
The form of the arguments of $\cG_n$ in eq.~\eqref{eq:Z4n_def} will be motivated in the next section.

As in the cubic case, we can rewrite the integrals over some of the differential forms above
in terms of incomplete elliptic integrals of first, second and third kind defined in eq.~\eqref{eq:FEPi}. 
Assuming the ordering $0<x<a_1<a_2<a_3<a_4<c$, 
it is easy to see that
\begin{alignat}{2}
&\Ef{0}{0}{x} =  &&- i\, \EF\!\left(\! \sqrt{\cra(x,a_1,a_4,a_2)} \big| 1-\lambda\right) - ( x \leftrightarrow 0 )\,,
\nonumber \\
&\Ef{-1}{c}{x} = && \frac{i\, y_c}{c_4 (c-a_4)} \left[ \EF\!\left(\! \sqrt{\cra(x,a_1,a_4,a_2)} \big| 1- \lambda \right) 
\right. \nonumber \\
&  && \left.+ \frac{a_{14}}{c-a_1} \Pi \left(\! \cra(a_2,a_1,a_4,c), \sqrt{\cra(x,a_1,a_4,a_2)}
\big| 1- \lambda \right) \right]  - ( x \leftrightarrow 0 )\,,\\
&\Ef{-1}{\infty}{x} = && -\frac{i}{c_4} \left[ a_4\,\EF\!\left(\! \sqrt{\cra(x,a_1,a_4,a_2)} \big| 1-\lambda \right) 
\right. \nonumber \\
\nonumber&  && \left.+ a_{14} \Pi \left(\! \cra(a_2,a_1,a_4,\infty), \sqrt{\cra(x,a_1,a_4,a_2)}
\big| 1-\lambda \right) \right]  - ( x \leftrightarrow 0 )\,,
\end{alignat}
where the cross ratio function cr is defined in eq.~\eqref{eq:lambda4}.
Similarly, we find, for $a_1<x<a_2$,
\begin{align}
Z_4(x) &= 2\,i\, \left [  \EE\!\left( \!\sqrt{ \cra(x,a_1,a_4,a_2) } \, \big| 1-\lambda \right) \right]
  + \frac{  y}{c_4(x-a_4)}\\ 
  \nonumber&
+ 2i\, \left[ 
   \frac{ (a_1+a_2)(a_3+a_4) -2 a_1 a_2 - 2 a_3 a_4 }{3\, a_{13}a_{24}} + 2 \frac{\eta_1}{ \omega_1}  \right] 
\EF\!\left(\!\sqrt{
\cra(x,a_1,a_4,a_2)}
\big| 1-\lambda \right) .
\end{align}
Similar relations exist for other regions. The signs in the previous relations
are related to our convention for the branches of the square root in
eq.~\eqref{eq:rsigns}.

Before we discuss in more detail the relationship between the iterated
integrals $\textrm{E}_4$ and the elliptic polylogarithms $\widetilde{\Gamma}$
in eq.~\eqref{eq:gamt_def}, let us compare the integration kernels in
eq.~\eqref{eq:final_quartic} to their cubic analogues in
eq.~\eqref{eq:final_Weierstrass}. First, just like in the cubic case,
eq.~\eqref{eq:E4_def} contains ordinary MPLs as a special case,
\beq\label{eq:E_to_G_4}
\Ef{1&\ldots&1}{c_1& \ldots& c_k}{x} = G(c_1,\ldots,c_k;x)\,,\qquad c_i\neq \infty\,.
\eeq
Second, we see that a main difference between the cubic case in eq.~\eqref{eq:final_Weierstrass} and the quartic case in eq.~\eqref{eq:final_quartic} is the appearance of the integration kernel $\psi_{-n}(\infty,x)$, which is absent in the cubic case. This is due to the fact that the point at infinity is a branch point in the cubic case, while it is not for a quartic polynomial. Conversely, we do not need to consider integration kernels of the form $\psi_{-n}(a_i,x)$, because they can always be removed using integration by parts. More generally, the counting of the integration kernels is identical in the cubic and quartic cases: for each $c\neq a_i$ there is a double infinite family of integration kernels, while for each branch point $a_i$ the infinite family $\psi_{-n}(a_i,x)$ is absent. Finally, another difference to the cubic case is the appearance of terms proportional to $\delta_{n2}$. These terms remove double poles at infinity that arise when multiplying $Z_4^{(1)}$ by another function with a simple pole at infinity. The role of these terms will become more transparent in the next section when discussing the connection between the iterated integrals $\textrm{E}_4$ and the elliptic polylogarithms $\widetilde{\Gamma}$.

Let us conclude by mentioning that the elliptic polylogarithms $\textrm{E}_4$ satisfy all the standard properties of iterated integrals. In particular, they form a shuffle algebra. In Appendix~\ref{app:shuffle_regularisation} we discuss regularisation procedure to regulate divergence at $x=0$ in eq.~\eqref{eq:E4_def} in a way that preserves the shuffle algebra structure.

\subsection{The relationship between $\textrm{E}_4$ and $\widetilde{\Gamma}$}
In this section we show that, just like in the cubic case, the elliptic polylogarithms $\textrm{E}_4$ can be written as linear combinations with constant complex coefficients of the iterated integrals $\widetilde{\Gamma}$. 

We start by showing that the differential one-forms $dx\,\psi_{\pm1}$ can be written as linear combinations of the one-forms $dz\,g^{(1)}$. We use the following identity relating the elliptic function $\kappa$ in eq.~\eqref{eq:kappa_def} and the function $g^{(1)}$,
\beq\label{eq:quartic_master_formula}
\frac{\kappa(z_c)}{\kappa(z)-\kappa(z_c)} = g^{(1)}(z_c-z_{\ast}) + g^{(1)}(z_c+z_{\ast})-g^{(1)}(z_c-z)-g^{(1)}(z_c+z)\,.
\eeq
This relation can be proved in the standard way\footnote{The argument presented here is strictly only valid if $z_c$ is not a half-period. It can, however, easily be extended to that case as well.}: Seen as a function of $z$, the left-hand side has simple poles only for $z=\pm z_{c}$, and the residues at these poles are $\pm1$. The right-hand side is periodic, despite individual terms not being periodic. Hence, the difference of the left- and right-hand sides defines an elliptic function without poles, and must thus be constant. The constant is zero, because both sides vanish for $z=z_{\ast}$. Using eq.~\eqref{eq:quartic_master_formula}, we immediately obtain for $c\neq\infty$,
\beq\bsp
dx\,\psi_1(c,x)&\,={dz}\,\frac{\kappa'(z)}{\kappa(z)-\kappa(z_c)}\\
&\,=dz\,\left[g^{(1)}(z-z_{c}) + g^{(1)}(z+z_{c})-g^{(1)}(z-z_{\ast})-g^{(1)}(z+z_{\ast})\right]\,,\\
dx\,\psi_{-1}(c,x)&\,={dz}\,\frac{\kappa'(z_c)}{\kappa(z)-\kappa(z_c)}\\
&\,=dz\,\left[g^{(1)}(z-z_{c}) - g^{(1)}(z+z_{c})+g^{(1)}(z_c-z_{\ast})+g^{(1)}(z_c+z_{\ast})\right]\,.
\esp\eeq
The corresponding relation for $dx\,\psi_{-1}(\infty,x)$ is given in eq.~\eqref{eq:kappa_g1}, while $dx\,\psi_{1}(\infty,x)$ requires one to know what $Z_4$ corresponds to when seen as a function on the torus. Using eq.~\eqref{eq:Phi4_wp} and the definition of the Weierstrass zeta function in eq.~\eqref{eq:zeta_def}, we find
\beq\label{eq:psi1_g1}
dx\,\psi_1(\infty,x)= -dz\,\left[g^{(1)}(z-z_{\ast}) +g^{(1)}(z+z_{\ast})\right]\,.
\eeq

Let us now analyse the one-forms $dx\,\psi_{\pm n}$ for $n>1$. Using eq.~\eqref{eq:quartic_master_formula} and~\eqref{eq:psi1_g1}, as well as the explicit definition of $\kappa$, we find
\beq\bsp
dx\,\psi_{n}(\infty,x) &\,= {dz}\,\cG_n\left(g^{(1)}(z);p_1(\kappa(z)),p_2(\kappa(z),c_{4}\,\kappa'(z)\right)\\
&\,= {dz}\,\cG_n\left(g^{(1)}(z);\wp(z),\wp'(z)\right)\\
&\,={dz}\,g^{(n)}(z)\,.
\esp\eeq
Since the remaining one-forms with $n>1$ can be obtained by multiplying the functions $\psi_{\pm1}$ by $Z^{(n-1)}_4(x)$, all other cases can easily be obtained by applying the Fay identity~\eqref{eq:Fay}, just like for the cubic case discussed in section~\ref{sec:torus_to_Weierstrass}. The only difference to the cubic case is the appearance of the terms proportional to Kronecker delta functions for $n=2$ in eq.~\eqref{eq:final_quartic}, which are required to cancel double poles at $z=\pm z_{\ast}$.
For example, we have
\beq\bsp
dx\,\psi_{-2}(\infty,x) &\,= \frac{x\,dx}{y}\,Z_4^{(1)}(x) - \frac{dx}{c_4}\\
&\,=-\,dz\,\kappa(z)\,\left[g^{(1)}(z-z_{\ast})+g^{(1)}(z+z_{\ast})\right]-dz\,\frac{\kappa'(z)}{c_4}\\
&\,=-\,dz\,\left(\frac{a_1}{c_4}+2g^{(1)}(z_{\ast})\right)\,\left[g^{(1)}(z-z_{\ast})+g^{(1)}(z+z_{\ast})\right]\\
&\,\phantom{=}-dz\,\left[g^{(1)}(z-z_{\ast})^2-g^{(1)}(z+z_{\ast})^2-\wp(z-z_{\ast})+\wp(z+z_{\ast})\right]\\
&\,=-\,dz\,\left(\frac{a_1}{c_4}+2g^{(1)}(z_{\ast})\right)\,\left[g^{(1)}(z-z_{\ast})+g^{(1)}(z+z_{\ast})\right]\\
&\,\phantom{=}-2dz\,\left[g^{(2)}(z-z_{\ast})-g^{(2)}(z+z_{\ast})\right]\,,
\esp\eeq
where in the last step we used eq.~\eqref{eq:g2g3_example}. We see that the double poles contained in the Weierstrass $\wp$ functions in eq.~\eqref{eq:g2g3_example} are precisely cancelled by the subtraction term $dx = dz\,\kappa'(z)$.

To conclude, we see that every one-form $dx\,\psi_{\pm n}$ can be written as a linear combination with constant complex coefficients of one-forms of the form $dz\,g^{(k)}(z\pm z_c)$. Hence, just like in the cubic case, every iterated integral $\textrm{E}_4$ can be written as a linear combination of elliptic polylogarithms $\widetilde{\Gamma}$.

\subsection{Algorithmic integration in the quartic case}
\label{sec:algorithm_4}

In this section we extend the integration algorithm of section~\ref{sec:algorithm} to the quartic case. More precisely, let us denote by $\cR_4\equiv \mathbb{C}(x,y)/\langle y^2=P_4(x)\rangle$ the field of rational functions of the elliptic curve $\cE$ defined by the quartic equation $y^2=P_4(x)$, and $\cA_4$ is the $\cR_4$-algebra generated by $Z_4(x)$ and all elliptic polylogarithms $\Ef{n_1&\ldots n_k}{c_1&\ldots c_k}{x}$,
\beq
\cA_4 = \Big\langle Z_4^{(m)}(x)\,\Ef{n_1&\ldots n_k}{c_1&\ldots c_k}{x}:m\ge 0, n_i\in\mathbb{Z}, c_i\in\widehat{\mathbb{C}}\Big\rangle_{\cR_4}\,.
\eeq
The total length is defined in the same way as in the cubic case, and it is easy to see that $\cA_4$ is filtered by the total length. In the following we present an algorithm which allows us to compute a primitive of every element in $\cA_4$. The algorithm is very similar to the cubic case, so we only highlight here the main similarities and differences.

We start by classifying integrals into the four classes defined in eq.~\eqref{eq:ABCD_classes}, and for each class we obtain recursion relations using integration by parts. The recursion relations for the types $A$ and $B$ have depth one, and we can reduce every integral in these types to a the integrals $A_{-1}$ and $B_{c,1}$, just like in the cubic case. For integrals of type $C$, however, the recursion relation has depth four in the quartic case (compared to depth three in the cubic case), and so every integral in this type can be written as a linear combination of $C_k$, $k\in\{-1,0,1,2\}$. Finally, integrals of type $D$ can be reduced to $D_{c,1}$ and integrals of type $C$, and we do not need to consider integrals of type $D$ with $c=a_i$. In the end, we find that every integral can be reduced to the following integrals
\begin{align}
\nonumber A_{-1}[\cX] &\,= \int \frac{dx}{x}\,\cX(x)\,,\qquad B_{c,1}[\cX] = \int \frac{dx}{x-c}\,\cX(x)\,,\qquad D_{c,1}[\cX] = \int \frac{dx}{y\,(x-c)}\,\cX(x)\,,\\
\label{eq:ABCD_MIs_2}C_{2}[\cX] &\,= \int \frac{x^2\,dx}{y}\,\cX(x)\,,\qquad C_{1}[\cX] = \int \frac{x\,dx}{y}\,\cX(x)\,,\\
\nonumber C_{0}[\cX] &\,= \int \frac{dx}{y}\,\cX(x)\,,\qquad C_{-1}[\cX] = \int \frac{dx}{y\,x}\,\cX(x)\,.
\end{align}
Comparing these integrals to eq.~\eqref{eq:ABCD_MIs}, we see that the only difference between the cubic and quartic cases is that we need to include $C_{2}[\cX]$ into the list of independent integrals. This reflects the fact that $x^2\,dx/y$ is related to the abelian differential of the second kind in the quartic case, cf.~eq.~\eqref{eq:tilde_Phi_4_def}.

The independent integrals in eq.~\eqref{eq:ABCD_MIs_2} are in one-to-one correspondance with the integration kernels in eq.~\eqref{eq:final_quartic}, with the exception of integrals of type $C_2$. 
Equivalently, we may consider the integrals
\beq
\widetilde{C}_2[Z^{(m)}_4\,\textrm{E}_4] = \int dx\,\Phi_4\,Z^{(m)}_4\,\textrm{E}_4\,.
\eeq
Integrating by parts, we obtain
\beq\label{eq:special_algo_0}
\widetilde{C}_2[Z^{(m)}_4\,\textrm{E}_4] = Z_4\,Z^{(m)}_4\,\textrm{E}_4 - \int dx\,Z_4\,\left(\textrm{E}_4\,\partial_xZ^{(m)}_4+ Z^{(m)}_4\,\partial_x\textrm{E}_4\right)\,.
\eeq
Just like in the cubic case, the integral in the right-hand side does not have lower total length, and so it cannot be done recursively. Indeed, eq.~\eqref{eq:universal_coefficient} and~\eqref{eq:Z4n_def} imply
\beq\label{eq:special_algo_1}
Z_4(z)\,\textrm{E}_4\,\partial_xZ^{(m)}_4 = \frac{m}{2}\,a_{12}\,a_{13}\,a_{14}\,\frac{1}{c_4\,y(x-a_1)}\,Z_4^{(m)}\,\textrm{E}_4+\ldots \,,
\eeq
Inserting eq.~\eqref{eq:special_algo_1} into eq.~\eqref{eq:special_algo_0}, we obtain an integral of type $D$ with $c=a_1$, and so this integral can be reduced to a linear combination of integrals of type $C$. We find
\beq\label{eq:special_algo_2}
\frac{m}{2c_4}\,a_{12}\,a_{13}\,a_{14}\,\int\frac{dx}{y(x-a_1)}\,Z_4^{(m)}\,\textrm{E}_4 = -m\int dx\,\Phi_4\,Z_4^{(m)}\,\textrm{E}_4 + \ldots\,,
\eeq
where the dots indicate terms that have lower total length, or they have the same total length and fall into the classes $C_0$ and $C_1$. Finally, inserting eq.~\eqref{eq:special_algo_2} into eq.~\eqref{eq:special_algo_1}, we have
\beq
(1+m)\int dx\,\Phi_4\,Z_4^{(m)}\,\textrm{E}_4 = \ldots\,,
\eeq
where the right-hand side only involves integrals with lower total length or from the class $C_0$ and $C_1$. All the integrals in the right-hand side have lower complexity and can be done recursively. This concludes the proof that every element in $\cA_4$ has a primitive that can be computed in an algorithmic way. Just like in the cubic case, this algorithm generalises the classical algorithm to evaluate integrals of rational functions on elliptic curves in terms of elliptic integrals.

%% file: examples.tex

\section{Examples}
\label{sec:examples}

In this section we discuss some examples of how to work with elliptic polylogarithms. The examples are of mostly mathematical nature. For an application to the sunrise integral, we refer to ref.~\cite{plumber_paper}.

\subsection{MPLs depending on square roots of cubic or quartic polynomials}
\label{sec:G_elliptic}
We have already seen that ordinary MPLs are special cases of elliptic polylogarithms, cf.~eqns.~\eqref{eq:E_to_G} and~\eqref{eq:E_to_G_4}. In this section we show that there are other classes of MPLs that can be expressed in terms of elliptic polylogarithms in a natural way.

Consider an MPL of the form $G(a_1(x,y),\ldots,a_n(x,y);a_{n+1}(x,y))$, where $a_i(x,y)$ is a rational function subject to the constraint $y^2=P_N(x)$, $N=3$ or $4$, In other words, we consider MPLs whose arguments are rational functions on the elliptic curve $\cE$ defined by the equation $y^2=P_N(x)$. In the following we only discuss the cubic case, $N=3$, and the quartic case is similar. Our goal is to show that every MPL of this type can be expressed in a natural way in terms of the elliptic polylogarithms $\textrm{E}_3$. The argument proceeds by induction in the weight $n$ of the MPL. We start by discussing the case $n=1$. Differentiating with respect to $x$, we find
\beq\label{eq:G_to_EG_1}
{\partial_x}G(a_1(x,y);a_{2}(x,y)) = \frac{1}{a_2(x,y)-a_1(x,y)}\,\left[a'_2(x,y) - \frac{a_2(x,y)}{a_1(x,y)}\,a'_1(x,y)\right]\,,
\eeq
where $a'_i(x,y) = \partial_xa_i(x,y) + \frac{1}{2y}\,P'_3(x)\,\partial_ya_i(x,y)$ is the (total) derivative  with respect to $x$. The right-hand side of eq.~\eqref{eq:G_to_EG_1} is obviously a rational function on $\cE$, and so its primitive can be expressed in terms of $Z_3(x)$ and $\Et{n}{c}{x}$. This concludes the proof that $G(a_1(x,y);a_{2}(x,y))$ can be expressed in terms of elliptic polylogarithms (and $Z_3$). The case of weight $n>1$ then follows by induction. Assume that the claim is true for MPLs up to weight $n-1$. We can then differentiate $G(a_1(x,y),\ldots,a_{n}(x,y);a_{n+1}(x,y))$ with respect to $x$, and since differentiation lowers the weight of MPLs, we know by induction that the derivative lies in $\cA_3$. Since every element in $\cA_3$ has a primitive in $\cA_3$, we conclude that $G(a_1(x,y),\ldots,a_{n}(x,y);a_{n+1}(x,y))$ lies in $\cA_3$, i.e., it can be expressed in terms of elliptic polylogarithms.

In the remainder of this section we illustrate the previous result on some simple examples. Let us consider the following function of weight one,
\beq
f(x) = \log\frac{1-y}{1+y} = \log\frac{1-\sqrt{P_3(x)}}{1+\sqrt{P_3(x)}}\,.
\eeq
Differentiating with respect to $x$, we find
\beq\label{eq:cubic_log_1}
\partial_xf(x) = \frac{3x^2-2s_1\,x+s_2}{y\,(P_3(x)-1)} = \frac{1}{y\,(x-b_1)}+\frac{1}{y\,(x-b_2)}+\frac{1}{y\,(x-b_3)}\,,
\eeq
where $b_i$ denote the roots of the cubic polynomial $P_3(x)-1$. It is possible to obtain explicit algebraic expressions for the roots $b_i$ in terms of the branch points $a_i$. The expressions are lengthy and not very illuminating for our purposes, so we do not show them here. We only mention the following useful relations,
\beq
y_{b_i} = \sqrt{P_3(b_i)} = 1 \textrm{~~~and~~~} b_1b_2b_3 = 1+a_1a_2a_3\,.
\eeq
We see from eq.~\eqref{eq:cubic_log_1} that the derivative of $f$ is a rational function on the elliptic curve $\cE$, and so $f$ itself can be written in terms of elliptic polylogarithms,
\beq
f(x) = c+\Et{-1}{b_1}{x}+\Et{-1}{b_2}{x}+\Et{-1}{b_1}{x}\,,
\eeq 
where $c$ may depend on the branch points $a_i$, but it is independent of $x$. The value of $c$ is determined from the fact that the elliptic polylogarithms vanish for $x=0$ while $f$ does not. We find
\beq\label{eq:cubic_log_2}
f(x) =\log\frac{1-y}{1+y}= \log\frac{1-y_0}{1+y_0}+\sum_{i=1}^3\Et{-1}{b_i}{x}\,.
\eeq
Before we turn to examples of higher weight, let us make some comments about the previous formula. We see that have managed to write an ordinary logarithm as a non-trivial linear combination of elliptic polylogarithms (non-trivial in the sense that the elliptic polylogarithms do not simply reduce to ordinary MPLs due to eq.~\eqref{eq:E_to_G}). Using eq.~\eqref{eq:E3_to_Pi}, we know that all the elliptic polylogarithms in the right-hand side of eq.~\eqref{eq:cubic_log_2} can be written as an incomplete ellipitic integral of the third kind, leading to an intriguing relation connecting a logarithm involving the square root of a cubic polynomial and incomplete elliptic integrals.

Alternatively, we also know that every $\textrm{E}_3$ function can be written as a linear combination of iterated integrals $\widetilde{\Gamma}$ on the torus. If $z_a$ is such that $(\mu(z_a),c_4\mu'(z_a)) = (a,y_a)$, eq.~\eqref{eq:E-1_example} gives
\begin{align}
\nonumber f(x) &\,= \log\frac{1-y_0}{1+y_0} + \sum_{i=1}^3\left[\gamt{1}{z_{b_i}}{z_x} - \gamt{1}{-z_{b_i}}{z_x} + 2\,g^{(1)}(z_{b_i})\,\gamt{0}{0}{z_x} - (z_x\leftrightarrow z_0)\right]\\
&\,= \sum_{i=1}^3\left[\gamt{1}{z_{b_i}}{z_x} - \gamt{1}{-z_{b_i}}{z_x} + 2\,g^{(1)}(z_{b_i})\,\gamt{0}{0}{z_x} - (z_x\leftrightarrow \omega_2/2)\right]\,,
\end{align}
where in the last step we used the fact that $f(a_i)=0$. Note that we can replace $\widetilde{\Gamma}$ by their periodic analogues $\Gamma$ in the previous equation, cf.~eq.~\eqref{eq:E3-1_to_Gamma}.

Finally, since $f(a_i)=0$, we see that eq.~\eqref{eq:cubic_log_2} implies a linear relation among $\Et{-1}{b_i}{a_j}$, $i,j\in\{1,2,3\}$. Similar relations of this type have appeared in ref.~\cite{Laporta:2004rb}, see eq.~(7.7) therein, where they have shown up in the context of the two-loop sunrise integral. While these relations were rather mysterious in ref.~\cite{Laporta:2004rb}, the analysis of this section reveals their origin: Ordinary logarithms with cubic or quartic roots inside their arguments can always be expressed in terms of elliptic polylogarithms. At some special points where the logarithm vanishes, this implies a linear relation among elliptic polylogarithms evaluated at those points. We stress that these special linear relations do not contradict the linear independence of the integration kernels $\varphi_{-1}(b_i,x)$. Indeed, the functions $\Et{-1}{b_i}{x}$ are linearly independent for generic values of $x$, and they are related only for certain special values of $x$. This effect is well-known in the context of hamonic polylogarithms: while all the functions $G(a_1,\ldots,a_n;x)$ with $a_i\in \{0,1\}$ are linearly independent for generic values of $x$, they reduce to multiple zeta values at the special point $x=1$, and there are additional relations among multiple zeta values, e.g., $G(1,0;1) = -G(0,1;1) = \zeta_2$.

Let us now illustrate how the previous story generalises to higher weights. As an example, we consider the functions
\beq
f_{\pm}(x) = \textrm{Li}_2\left(\frac{1-y}{2}\right) \pm \textrm{Li}_2\left(\frac{1+y}{2}\right)= \textrm{Li}_2\left(\frac{1-\sqrt{P_3(x)}}{2}\right) \pm \textrm{Li}_2\left(\frac{1+\sqrt{P_3(x)}}{2}\right)\,.
\eeq
The decomposition into even and odd parts is not essential to the discussion, but it makes some of the formulas more compact. We only discuss the odd combination $f_-$ in detail, because the even case is very similar. Differentiating with respect to $x$, we find
\begin{align}
\partial_x&f_-(x)= \frac{1}{2}\left[\frac{1}{y}\log\frac{1+a_1a_2a_3}{4}-\log\frac{1-y}{1+y}+\frac{1}{y}\sum_{i=1}^3\log\left(1-\frac{x}{b_i}\right)
\right]\!\sum_{i=1}^3\frac{1}{x-b_i}\\
\nonumber\!\!&\!\!\!\,= \frac{1}{2}\left[\frac{1}{y}\log\frac{1+a_1a_2a_3}{4}-\log\frac{1-y_0}{1+y_0}+\sum_{i=1}^3\left(\frac{1}{y}\Et{1}{b_i}{x}-\Et{-1}{b_i}{x}\!\right)
\right]\!\sum_{i=1}^3\frac{1}{x-b_i}\,,
\end{align}
where in the last step we have used eqns.~\eqref{eq:E_to_G} and~\eqref{eq:cubic_log_2}. As expected, the derivative takes values in the algebra $\cA_3$, and so it admits a primitive inside the same space, i.e., $f_-$ can be expressed in terms of elliptic polylogarithms. We find
\begin{align}\label{eq:f_minus}
f_-(x)&\, = \textrm{Li}_2\left(\frac{1-y_0}{2}\right) - \textrm{Li}_2\left(\frac{1+y_0}{2}\right)-\frac{1}{2}\log\frac{1-y_0}{1+y_0}\,\sum_{i=1}^3\log\left(1-\frac{x}{b_i}\right) \\
\nonumber&\,+\frac{1}{2}\log\frac{1+a_1a_2a_3}{4}\sum_{i=1}^3\Et{-1}{b_i}{x}+ \frac{1}{2}\sum_{i,j=1}^3\left(\Et{-1&1}{b_i&b_j}{x}-\Et{1&-1}{b_i&b_j}{x}\right)\,.
\end{align}
Similarly, we find
\begin{align}\label{eq:f_plus}
f_+(x)&\, = \textrm{Li}_2\left(\frac{1-y_0}{2}\right) + \textrm{Li}_2\left(\frac{1+y_0}{2}\right)+\frac{1}{2}\log\frac{1-y_0}{1+y_0}\,\sum_{i=1}^3\log\left(1-\frac{x}{b_i}\right) \\
\nonumber&\,-\frac{1}{2}\log\frac{1+a_1a_2a_3}{4}\sum_{i=1}^3\Et{-1}{b_i}{x}+ \frac{1}{2}\sum_{i,j=1}^3\left(\Et{-1&-1}{b_i&b_j}{x}-G(b_i,b_j;x)\right)\,.
\end{align}
We see that $f_{\pm}$ can always be expressed in terms of elliptic polylogarithms and ordinary MPLs. Using the results of section~\ref{sec:torus_to_Weierstrass} we can also write the previous relations in terms of the functions $\widetilde{\Gamma}$ or $\Gamma$.

Let us conclude this section with some comments about the relevance of the results in this section for the computation of Feynman integrals. The results of this section show that it is possible to find combinations of genuine elliptic polylogarithms that evaluate to ordinary MPLs with square-root arguments. These examples show that the distinction between ordinary and elliptic MPLs may not be as clear-cut as sometimes assumed in the physics literature, and some care may be needed when making statements about when a given Feynman integral or amplitude can or cannot be expressed in terms of ordinary MPLs alone. For example, it could be conceivable that an amplitude is expressed as a sum of integrals which individually evaluate to elliptic polylogarithms that cannot be written as ordinary MPLs, but their sum only involves MPLs, through a formula similar to eq.~\eqref{eq:f_minus} or~\eqref{eq:f_plus}.


\subsection{Hypergeometric ${}_2F_1$ functions that evaluate to elliptic polylogarithms}
\label{sec:2F1}

Consider the following class of integrals
\beq\label{eq:T_def}
T(n_1,n_3,n_3;z) = \int_0^1dx\,x^{-1/2+n_1+\alpha_1\eps}\,(1-x)^{-1/2+n_2+\alpha_2\eps}\,(1-zx)^{-1/2+n_3+\alpha_3\eps}\,,
\eeq
where $n_i$ and $\alpha_i$ are integers, and for concreteness we assume $0<z<1$. This class of integrals is tightly connected to Gauss' hypergeometric function,
\beq
{}_2F_1(a,b,c;z) = \frac{\Gamma(c)}{\Gamma(c-b)\,\Gamma(b)}\int_0^1dx\,x^{b-1}\,(1-x)^{c-b-1}\,(1-zx)^{-a}\,.
\eeq

We are interested in the Laurent expansion in $\eps$ of the integral. In the case where the exponents in the integrand are integers for $\eps=0$, the Laurent coefficients are MPLs that can be computed explicitly using standard techniques. In the case where all the exponents are half-integers, however, the Laurent coefficients are not known in the literature. In the remainder of this section we show that the Laurent coefficients of $T$ can be expressed in terms of elliptic polylogarithms $\textrm{E}_3$.

Using integration by parts, one can show that every integral in the family defined by eq.~\eqref{eq:T_def} can be written as a linear combination of two master integrals. We choose the following basis integrals,
\begin{align}
T_1(z) &\,= T(0,0,0;z) = \frac{1}{\sqrt{z}}\int_0^1\frac{dx}{y}\,x^{\alpha_1\eps}\,(1-x)^{\alpha_2\eps}\,(1-zx)^{\alpha_3\eps}\,,\\
\nonumber T_2(z) &\,= \frac{1+\lambda}{3}\,T(0,0,0;z)-T(1,0,0;z) = \frac{1}{2z}\int_0^1{dx}\,\widetilde{\Phi}_3(x)\,x^{\alpha_1\eps}\,(1-x)^{\alpha_2\eps}\,(1-zx)^{\alpha_3\eps}\,,
\end{align}
with $\lambda\equiv 1/z$ and $y^2=x(x-1)(x-\lambda)$.

Let us discuss the computation of $T_1$. After expansion in $\eps$, the
integrand involves powers of logarithms that can be recast in the form of MPLs.
Using eq.~\eqref{eq:E_to_G}, we see that all integrals can be reduced to
integrals of the type
\beq
\int_0^1\frac{dx}{y}\,G(c_1,\ldots,c_k;x) = \int_0^1\frac{dx}{y}\,\Et{1&\ldots&1}{c_1&\ldots&c_k}{x} =  2\sqrt{z}\,\Et{0&1&\ldots&1}{0&c_1&\ldots&c_k}{1}\,,
\eeq
with $c_i\in\{0,1,\lambda\}$. For the first few orders, we find explicitly
\beq\bsp
\label{eq:2F1_T1}
T_1(z) &\,= 2\,\Et{0}{0}{1}+ 2\,{\eps}\,\left[\alpha_1\,\Et{0&1}{0&0}{1}+\alpha_2\,\Et{0&1}{0&1}{1} + \alpha_3\,\Et{0&1}{0&\lambda}{1}\right] \\
&\,+ 
2\,{\eps^2}\,\Big[\alpha_1^2\,\Et{0&1&1}{0&0&0}{1} + \alpha_1\alpha_2\,\left(\Et{0&1&1}{0&0&1}{1}+\Et{0&1&1}{0&1&0}{1}\right) +\alpha_2^2\,\Et{0&1&1}{0&1&1}{1}\\
&\,+ \alpha_1\alpha_3\,\left(\Et{0&1&1}{0&0&\lambda}{1}+\Et{0&1&1}{0&\lambda&0}{1}\right)  + \alpha_2\alpha_3\,\left(\Et{0&1&1}{0&1&\lambda}{1}+\Et{0&1&1}{0&\lambda&1}{1}\right)\\
&\, + \alpha_3^2\,\Et{0&1&1}{0&\lambda&\lambda}{1}\Big]+\ord(\eps^3)\,.
\esp\eeq 
We see that eq.~\eqref{eq:2F1_T1} involves at every order in $\eps$ only functions of uniform weight. More precisely, all the terms in the coefficient of $\eps^k$ have weight $k$ (we recall that the weight of $\Et{n_1&\ldots&n_k}{c_1&\ldots&c_k}{x}$ is not just the number of integrations, but it is defined as $|n_1|+\ldots+|n_k|$). 

Let us now discuss the computation of the integral $T_2$. Using the algorithm described in section~\ref{sec:algorithm}, we find 
\beq\label{eq:T2_to_T2bar}
T_2(z) = \frac{1}{(1+2(\alpha_1+\alpha_2+\alpha_3)\eps)\,z}\,\left[\frac{2\eta_1}{\omega_1}\,T_1(z) + \overline{T}_2(z)\right]\,,
\eeq
with
\begin{align}
\nonumber\overline{T}_2(z) &\,=  {\eps}\Big[2(\alpha_1+\alpha_2+\alpha_3)\Et{2}{\infty}{1}-\alpha_3\,\frac{2\pi i}{\omega_1}\,\Et{1}{\lambda}{1}-\frac{\alpha_1}{2}\Et{2}{0}{1}-\frac{\alpha_2}{2}\,\ERegt{2}{1}\\
\label{eq:T2bar}&\,-\frac{\alpha_3}{2}\,\Et{2}{\lambda}{1}\Big]
+{\eps}^2\Big[2(\alpha_1+\alpha_2+\alpha_3)\Big(\alpha_1\,\Et{2&1}{\infty&0}{1}+\alpha_2\,\Et{2&1}{\infty&1}{1}\\
\nonumber&\,+\alpha_3\,\Et{2&1}{\infty&\lambda}{1}\Big)+\frac{2\pi i}{\omega_1}\Big(\alpha_1\alpha_2\,\Et{1&1}{0&1}{1}-\alpha_1\alpha_3\,\Et{1&1}{\lambda&0}{1}-\alpha_3^2\,\Et{1&1}{\lambda&\lambda}{1}\Big)\\
\nonumber&\,-\frac{\alpha_1^2}{2}\,\Et{2&1}{0&0}{1}-\frac{\alpha_2^2}{2}\,\ERegt{2&1}{1&1}-\frac{\alpha_3^2}{2}\,\Et{2&1}{\lambda&\lambda}{1}-\frac{\alpha_1\alpha_2}{2}\,\Big(\ERegt{2}{1}\,\Et{1}{0}{1}-\Et{1&2}{0&1}{1}\\
\nonumber&\,+\Et{2&1}{0&1}{1}\Big)-\frac{\alpha_1\alpha_3}{2}\,\Big(\Et{2&1}{0&\lambda}{1}+\Et{2&1}{\lambda&0}{1})\Big)
-\frac{\alpha_2\alpha_3}{2}\,\Big(\ERegt{2}{1}\,\Et{1}{\lambda}{1}\\
\nonumber&\,-\Et{1&2}{\lambda&1}{1}+\Et{2&1}{\lambda&1}{1}\Big)\Big]+\ord(\eps^3)\,,
\end{align}
where the $\varepsilon_3$ denote regularised values at the upper integration limit. These arise because individual terms may diverge logarithmically at 1, e.g., if $\delta\to 0$, we have
\beq\bsp\label{eq:reg_values}
\Et{1}{1}{1-\delta} &\,= \log\delta\,,\\
\Et{2}{1}{1-\delta}&\, = Z_3(1)\,\log\delta+\ERegt{2}{1}+\ord(\delta) = -\frac{4\pi i}{\omega_1}\,\log\delta+\ERegt{2}{1}+\ord(\delta)\,,\\
\Et{2&1}{1&1}{1-\delta}&\, = Z_3(1)\,\frac{1}{2}\log^2\delta+\ERegt{2&1}{1&1}+\ord(\delta)=-\frac{2i\pi}{\omega_1}\log^2\delta+\ERegt{2&1}{1&1}+\ord(\delta)\,.
\esp\eeq
The finite terms are explicitly given by
\beq\bsp\label{eq:reg_values_2}
\ERegt{2}{1} &\,= \int_0^1dx\,\left[\varphi_{2}(x)-Z_3(1)\,\varphi_1(x)\right]\\
&\,= \int_0^1{dx}\,\left[\frac{Z_3(x)}{x-1}+\frac{Z_3(x)^2{\sqrt{\lambda}}}{2y}+\frac{4\pi i}{\omega_1(x-1)}\right]\,,\\
\ERegt{2&1}{1&1} &\,= \int_0^1dx\,\left[\varphi_{2}(x)-Z_3(1)\,\varphi_1(x)\right]\Et{1}{1}{x}\\
&\,= \int_0^1{dx}\,\log(1-x)\left[\frac{Z_3(x)}{x-1}+\frac{Z_3(x)^2{\sqrt{\lambda}}}{2y}+\frac{4\pi i}{\omega_1(x-1)}\right]\,.
\esp\eeq
The logarithmic singularities all cancel in the final expression for $\overline{T}_2$, leaving a finite result. More details on the regularisation can be found in appendix~\ref{app:shuffle_regularisation}.

Let us make some comments about eq.~\eqref{eq:T2bar}. First, we see that the result involves elliptic polylogarithms of the form $\Et{2&\ldots}{\ast &\ldots}{x}$. We thus see the necessity for the integration kernels $\varphi_{\pm n}$ with $n>1$.
Second, let us comment on the weight of the Laurent coefficients in eq.~\eqref{eq:T2bar}. Since $\omega_1=\Et{0}{0}{1}$ has weight zero, we see that all the terms in eq.~\eqref{eq:T2bar} have uniform weight\footnote{The weight of $\ERegt{n_1&\ldots&n_k}{c_1&\ldots&c_k}$ is defined to be identical to the weight of $\Et{n_1&\ldots&n_k}{c_1&\ldots&c_k}{x}$.}. We have checked that this observation remains true for the first six terms in the Laurent expansion. Furthermore, since the periods $\omega_i$ have weight zero, the Legendre relation~\eqref{eq:Legendre} implies that we should assign weight one to the quasi-periods $\eta_i$ (and thus also to $Z_3$). We observe that with this assignment of the weight, all the terms in the right-hand side of eq.~\eqref{eq:T2_to_T2bar} have uniform weight once the overall prefactor is scaled out. We emphasise that in this context it is important that the concept of weight is \emph{not} associated to the number of integrations!

\subsection{Appell $F_1$ functions that evaluate to elliptic polylogarithms}
In this section we study two different classes of Appell $F_1$ functions that evaluate to elliptic polylogarithms. The Appell $F_1$ function admits the integral representation
\beq\label{eq:AppellF1}
F_1(a,b_1,b_2,c;z_1,z_2) = \frac{\Gamma(c)}{\Gamma(c-a)\,\Gamma(a)}\int_0^1dx\,x^{a-1}\,(1-x)^{c-a-1}\,(1-z_1x)^{-b_1}\,\,(1-z_2x)^{-b_2}\,.
\eeq
We assume that $(a,b_1,b_2,c)$, and thus the exponents in the integrand, depend linearly on $\eps$. In the following we show that in the case where for $\eps=0$ three or more of the exponents are half-integers, the coefficients appearing in the Laurent expansion around $\eps=0$ can be expressed in terms of elliptic polylogarithms.

\subsubsection{The cubic case}
We consider the following family of integrals, 
\beq\bsp
A&(n_1,n_3,n_3,n_4;z_1,z_2)\\
&\, = \int_0^1dx\,x^{-1/2+n_1+\alpha_1\eps}\,(1-x)^{-1/2+n_2+\alpha_2\eps}\,(1-z_1x)^{-1/2+n_3+\alpha_3\eps}\,(1-z_2x)^{n_4+\alpha_4\eps}\,,
\esp\eeq
where $n_i$ and $\alpha_i$ are integers, and for concreteness we assume $0<z_i<1$. These integrals are closely related to the Appell $F_1$ function defined in eq.~\eqref{eq:AppellF1}. Using integration by parts, we can write every integral in this family as a linear combination of the following three master integrals
\begin{align}
\nonumber A_1(z_1,z_2) &\,= A(0,0,0,0;z_1,z_2)\\
\nonumber&\, = \frac{1}{\sqrt{z_1}}\int_0^1\frac{dx}{y}\,x^{\alpha_1\eps}\,(1-x)^{\alpha_2\eps}\,(1-z_1x)^{\alpha_3\eps}\,(1-z_2x)^{\alpha_4\eps}\,,\\
 A_2(z_1,z_2) &\,= \frac{1+\lambda_1}{3}\,A(0,0,0,0;z_1,z_2)-A(1,0,0,0;z_1,z_2)\\
\nonumber&\, = \frac{1}{2{z_1}}\int_0^1{dx}\,\widetilde{\Phi}_3(x)\,x^{\alpha_1\eps}\,(1-x)^{\alpha_2\eps}\,(1-z_1x)^{\alpha_3\eps}\,(1-z_2x)^{\alpha_4\eps}\,,\\
\nonumber A_3(z_1,z_2) &\,= A(0,0,0,-1;z_1,z_2)\\
\nonumber&\, = \frac{1}{z_2\sqrt{z_1}}\int_0^1\frac{dx}{y(x-\lambda_2)}\,x^{\alpha_1\eps}\,(1-x)^{\alpha_2\eps}\,(1-z_1x)^{\alpha_3\eps}\,(1-z_2x)^{\alpha_4\eps}\,,
\end{align}
with $\lambda_i\equiv 1/z_i$ and $y^2=x(x-1)(x-\lambda_1)$.

The computation of $A_1$ and $A_2$ is completely analogous to the case of the ${}_2F_1$ function. For the first master integral we find,
\begin{align}
\nonumber A_1(z_1,z_2) &\,= T_1(z_1)+ 2\,{\eps}\,\alpha_4\,\Et{0&1}{0&\lambda_2}{1} + 
2\,{\eps^2}\,\alpha_4\Big[
\alpha_1\,\left(\Et{0&1&1}{0&0&\lambda_2}{1}+\Et{0&1&1}{0&\lambda_2&0}{1}\right)\\
\nonumber&\,
+ \alpha_2\,\left(\Et{0&1&1}{0&1&\lambda_2}{1}+\Et{0&1&1}{0&\lambda_2&1}{1}\right)
+ \alpha_3\,\left(\Et{0&1&1}{0&\lambda_1&\lambda_2}{1}+\Et{0&1&1}{0&\lambda_2&\lambda_1}{1}\right)\\
&\,+ \alpha_4\,\Et{0&1&1}{0&\lambda_2&\lambda_2}{1}\Big]+\ord(\eps^3)\,,
\end{align} 
where $T_1$ is given in eq.~\eqref{eq:2F1_T1}.
The second master integral is given by
\beq\label{eq:A2_to_A2bar}
A_2(z_1,z_2) = \frac{1}{(1+2(\alpha_1+\alpha_2+\alpha_3+\alpha_4)\eps)\,z_1}\,\left[\frac{2\eta_1}{\omega_1}\,A_1(z_1,z_2) + \overline{A}_2(z_1,z_2)\right]\,,
\eeq
with
\begin{align}
 \overline{A}_2&(z_1,z_2) =  \overline{T}_2(z_1)+{\eps}\,\alpha_4\,\Big[2\,\Et{2}{\infty}{1}-\frac{2\pi i}{\omega_1}\,\Et{1}{\lambda_2}{1}-\frac{1}{2}\,\Et{2}{\lambda_2}{1}\Big]\\
\nonumber&\,+{\eps}^2\, \alpha_4\, \Big[-\frac{2\pi i}{\omega_1}\Big(\alpha_1\,\Et{1&1}{\lambda_2&0}{1}+\alpha_3\,\left( \Et{1&1}{\lambda_1&\lambda_2}{1}+\,\Et{1&1}{\lambda_2&\lambda_1}{1} \right)+\alpha_4\,\Et{1&1}{\lambda_2&\lambda_2}{1}\Big)\\
\nonumber&\,+\alpha_1\,\Big(2\,\Et{2&1}{\infty&\lambda_2}{1}+2\,\Et{2&1}{\infty&0}{1}-\frac{1}{2}\,\Et{2&1}{0&\lambda_2}{1}-\frac{1}{2}\,\Et{2&1}{\lambda_2&0}{1}\Big)\\
\nonumber&\,+\alpha_2\,\Big(2\,\Et{2&1}{\infty&\lambda_2}{1}+2\,\Et{2&1}{\infty&1}{1}-\frac{1}{2}\,\ERegt{2}{1}\,\Et{1}{\lambda_2}{1}-\frac{1}{2}\,\Et{2&1}{\lambda_2&1}{1}\\
\nonumber&\,+\frac{1}{2}\,\Et{1&2}{\lambda_2&1}{1}\Big)
+\alpha_3\,\Big(2\,\Et{2&1}{\infty&\lambda_2}{1}+2\,\Et{2&1}{\infty&\lambda_1}{1}-\frac{1}{2}\,\Et{2&1}{\lambda_1&\lambda_2}{1}\\
\nonumber&\,-\frac{1}{2}\,\Et{2&1}{\lambda_2&\lambda_1}{1}\Big)
+\alpha_4\,\Big(2\,\Et{2&1}{\infty&\lambda_2}{1}-\frac{1}{2}\,\Et{2&1}{\lambda_2&\lambda_2}{1}\Big)
\Big]+\ord(\eps^3)\,.
\end{align}
The third master integral has an additional logarithmic singularity at $x=\lambda_2$, and all the integrations can be performed using the formula
\beq
\int_0^1\frac{dx}{y(x-\lambda_2)}\,G(c_1,\ldots,c_k;x) =  \frac{1}{y_{\lambda_2}}\Et{-1&1&\ldots&1}{\lambda_2&c_1&\ldots&c_k}{1}\,.
\eeq
We find
\begin{equation}
{A}_3(z_1,z_2) = -\frac{1}{z_2\sqrt{z_1}y_{\lambda_2}}\, \overline{A}_3(z_1,z_2)\,,
\end{equation}
\begin{align}
 \overline{A}_3&(z_1,z_2) = \Et{-1}{\lambda_2}{1} + \eps\, \Big[
 \alpha_1\,\Et{-1&1}{\lambda_2&0}{1}+ \alpha_2\,\Et{-1&1}{\lambda_2&1}{1} \\
 \nonumber&\,+  \alpha_3\,\Et{-1&1}{\lambda_2&\lambda_1}{1}+ \alpha_4\,\Et{-1&1}{\lambda_2&\lambda_2}{1}
\Big]
+ \eps^2 \, \Big[
 \alpha_1^2\,\Et{-1&1&1}{\lambda_2&0&0}{1}\\
 \nonumber&\,+\alpha_1\alpha_2\Big(\Et{-1&1&1}{\lambda_2&0&1}{1}+\Et{-1&1&1}{\lambda_2&1&0}{1}\Big)
 +\alpha_1\alpha_3\Big(\Et{-1&1&1}{\lambda_2&0&\lambda_1}{1}+\Et{-1&1&1}{\lambda_2&\lambda_1&0}{1}\Big)\\
\nonumber &\,  +\alpha_1\alpha_4\Big(\Et{-1&1&1}{\lambda_2&0&\lambda_2}{1}+\Et{-1&1&1}{\lambda_2&\lambda_2&0}{1}\Big)
+\alpha_2^2\,\Et{-1&1&1}{\lambda_2&1&1}{1}\\
\nonumber &\, +\alpha_2\alpha_3\Big(\Et{-1&1&1}{\lambda_2&1&\lambda_1}{1}+\Et{-1&1&1}{\lambda_2&\lambda_1&1}{1}\Big)
+\alpha_2\alpha_4\Big(\Et{-1&1&1}{\lambda_2&1&\lambda_2}{1}+\Et{-1&1&1}{\lambda_2&\lambda_2&1}{1}\Big)\\
\nonumber&\,+\alpha_3^2\,\Et{-1&1&1}{\lambda_2&\lambda_1&\lambda_1}{1}
+\alpha_3\alpha_4\Big(\Et{-1&1&1}{\lambda_2&\lambda_1&\lambda_2}{1}+\Et{-1&1&1}{\lambda_2&\lambda_2&\lambda_1}{1}\Big)\\
\nonumber&\,+\alpha_4^2\,\Et{-1&1&1}{\lambda_2&\lambda_2&\lambda_2}{1}\Big]+\ord(\eps^3)\,.
\end{align}
We observe that, just like in the case of the ${}_2F_1$ function, the Laurent coefficients of the master integrals $A_i$ have uniform weight.

\subsubsection{The quartic case}
We now consider the following family of integrals, which corresponds to the case where all the exponents in the integrand in eq.~\eqref{eq:AppellF1} evaluate to half-integers for $\eps=0$,
\begin{align}
B&(n_1,n_3,n_3,n_4;z_1,z_2)\\
\nonumber&\, = \int_0^1dx\,x^{-1/2+n_1+\alpha_1\eps}\,(1-x)^{-1/2+n_2+\alpha_2\eps}\,(1-z_1x)^{-1/2+n_3+\alpha_3\eps}\,(1-z_2x)^{-1/2+n_4+\alpha_4\eps}\,,
\end{align}
where $n_i$ and $\alpha_i$ are integers, and we assume $0<z_2<z_1<1$. Just like in the cubic case, there are three master integrals for this family, which we choose as
\begin{align}
\nonumber B_1(z_1,z_2) &\,= B(0,0,0,0;z_1,z_2)\\
\nonumber&\, = -\frac{i}{\sqrt{z_1z_2}}\int_0^1\frac{dx}{y}\,x^{\alpha_1\eps}\,(1-x)^{\alpha_2\eps}\,(1-z_1x)^{\alpha_3\eps}\,(1-z_2x)^{\alpha_4\eps}\,,\\
 B_2(z_1,z_2) &\,= B(2,0,0,0;z_1,z_2)-\frac{1+\lambda_1+\lambda_2}{2}\,B(1,0,0,0;z_1,z_2)\\
 \nonumber&\,\phantom{=}+\frac{\lambda_1\lambda_2+\lambda_1+\lambda_2}{6}B(0,0,0,0;z_1,z_2)\\
\nonumber&\, = -\frac{i\sqrt{1-z_2}}{2z_1z_2}\int_0^1{dx}\,\widetilde{\Phi}_4(x)\,x^{\alpha_1\eps}\,(1-x)^{\alpha_2\eps}\,(1-z_1x)^{\alpha_3\eps}\,(1-z_2x)^{\alpha_4\eps}\,,\\
\nonumber B_3(z_1,z_2) &\,= B(1,0,0,0;z_1,z_2)\\
\nonumber&\, = -\frac{i}{\sqrt{z_1z_2}}\int_0^1\frac{x\,dx}{y}\,x^{\alpha_1\eps}\,(1-x)^{\alpha_2\eps}\,(1-z_1x)^{\alpha_3\eps}\,(1-z_2x)^{\alpha_4\eps}\,,
\end{align}
with $\lambda_i\equiv 1/z_i$ and $y^2=x(x-1)(x-\lambda_1)(x-\lambda_2)$. Since $\lambda_2>\lambda_1>1$, our convention in eq.~\eqref{eq:rsigns} implies $\textrm{Im }y<0$ for $0<x<1$. All the integrations can easily be done order by order in $\eps$ in terms of elliptic polylogarithms. For the first master integral, we find
\begin{equation}
B_1(z_1,z_2) = \frac{2}{i \sqrt{1-z_2}} \overline{B}_1(z_1,z_2)\,,
\end{equation}
with
\begin{align}
\nonumber \overline{B}&_1(z_1,z_2) =\,\Ef{0}{0}{1} + \eps\,\Big[\alpha_1\,\Ef{0&1}{0&0}{1} + \alpha_2\,\Ef{0&1}{0&1}{1} 
+\alpha_3\,\Ef{0&1}{0&\lambda_1}{1}\\
\nonumber&\,+\alpha_4\,\Ef{0&1}{0&\lambda_2}{1}\Big] + \eps^2\,\Big[\alpha_1^2\,\Ef{0&1&1}{0&0&0}{1}
+\alpha_1\alpha_2\,\Big(\Ef{0&1&1}{0&0&1}{1}+\Ef{0&1&1}{0&1&0}{1}\Big)\\
&\,+\alpha_1\alpha_3\,\Big(\Ef{0&1&1}{0&0&\lambda_1}{1}+\Ef{0&1&1}{0&\lambda_1&0}{1}\Big)
+\alpha_1\alpha_4\,\Big(\Ef{0&1&1}{0&0&\lambda_2}{1}+\Ef{0&1&1}{0&\lambda_2&0}{1}\Big)\\
\nonumber&\,+\alpha_2^2\,\Ef{0&1&1}{0&1&1}{1}
+\alpha_2\alpha_3\,\Big(\Ef{0&1&1}{0&1&\lambda_1}{1}+\Ef{0&1&1}{0&\lambda_1&1}{1}\Big)+\alpha_2\alpha_4\,\Big(\Ef{0&1&1}{0&1&\lambda_2}{1}\\
\nonumber&\,+\Ef{0&1&1}{0&\lambda_2&1}{1}\Big)
+\alpha_3^2\,\Ef{0&1&1}{0&\lambda_1&\lambda_1}{1}+\alpha_3\alpha_4\,\Big(\Ef{0&1&1}{0&\lambda_1&\lambda_2}{1}+\Ef{0&1&1}{0&\lambda_2&\lambda_1}{1}\Big)\\
\nonumber&\,+\alpha_4^2\,\Ef{0&1&1}{0&\lambda_2&\lambda_2}{1}\Big]+\ord(\eps^3)\,.
\end{align}
For the second master integral, we find
\beq
B_2(z_1,z_2) = \frac{i\sqrt{1-z_2}}{2z_1z_2}\,\left[\frac{4\,\eta_1}{\omega_1}\,  \overline{B}_1(z_1,z_2) + \frac{1}{1+(\alpha_1+\alpha_2+\alpha_3+\alpha_4)\eps}\,\overline{B}_2(z_1,z_2)\right]\,,
\eeq
with
\begin{align}
\overline{B}_2&(z_1,z_2) = -\frac{2\pi i}{\omega_1}+\eps\,\Big[\alpha_1\,\Ef{2}{0}{1} + \alpha_2\,\ERegf{2}{1} + \alpha_3\,\Ef{2}{\lambda_1}{1}+ \alpha_4\,\Ef{2}{\lambda_2}{1} \\
\nonumber&\,-\frac{2\pi i}{\omega_1}\Big(\alpha_1\,\Ef{1}{0}{1}+\alpha_3\,\Ef{1}{\lambda_1}{1}+\alpha_4\,\Ef{1}{\lambda_2}{1}\Big)\Big] +\eps^2\Big[\alpha_1^2\,\Ef{2&1}{0&0}{1} 
\\
\nonumber&\,+\alpha_1\alpha_2\Big(\ERegf{2}{1}\Ef{1}{0}{1}  - \Ef{1&2}{0&1}{1} + \Ef{2&1}{0&1}{1}\Big)+\alpha_1\alpha_3\Big(\Ef{2&1}{0&\lambda_1}{1} + \Ef{2&1}{\lambda_1&0}{1}\Big)\\
\nonumber&\,+\alpha_1\alpha_4\Big(\Ef{2&1}{0&\lambda_2}{1} + \Ef{2&1}{\lambda_2&0}{1} \Big)+\alpha_2^2\,\ERegf{2&1}{1&1} 
+\alpha_2\alpha_3\Big(\ERegf{2}{1}\Ef{1}{\lambda_1}{1} - \Ef{1&2}{\lambda_1&1}{1} \\
\nonumber&\,+ \Ef{2&1}{\lambda_1&1}{1} \Big)
+\alpha_2\alpha_4\Big(\ERegf{2}{1}\Ef{1}{\lambda_2}{1} - \Ef{1&2}{\lambda_2&1}{1} + \Ef{2&1}{\lambda_2&1}{1} \Big)
+\alpha_3^2\,\Ef{2&1}{\lambda_1&\lambda_1}{1} \\
\nonumber&\,+\alpha_3\alpha_4\Big(\Ef{2&1}{\lambda_1&\lambda_2}{1}+\Ef{2&1}{\lambda_2&\lambda_1}{1} \Big)+\alpha_4^2\,\Ef{2&1}{\lambda_2&\lambda_2}{1} 
-\frac{2i\pi}{\omega_1}\Big(\alpha_1^2\,\Ef{1&1}{0&0}{1}\\
\nonumber&\,+\alpha_1\alpha_3\,\Big(\Ef{1&1}{0&\lambda_1}{1}+\Ef{1&1}{\lambda_1&0}{1}\Big)+\alpha_1\alpha_4\,\Big(\Ef{1&1}{0&\lambda_2}{1}+\Ef{1&1}{\lambda_2&0}{1}\Big)\\
\nonumber&\,
+\alpha_3^2\,\Ef{1&1}{\lambda_1&\lambda_1}{1}+\alpha_3\alpha_4\,\Big(\Ef{1&1}{\lambda_1&\lambda_2}{1}+\Ef{1&1}{\lambda_2&\lambda_1}{1}\Big)+\alpha_4^2\,\Ef{1&1}{\lambda_2&\lambda_2}{1}\Big)
\Big]+\ord(\eps^3)\,.
\end{align}
Finally, the third master, which is related to an abelian differential with a simple pole at infinity, is given by
\begin{equation}
B_3(z_1,z_2) = \frac{1}{ i\, \sqrt{z_1z_2}}\, \overline{B}_3(z_1,z_2)\,,
\end{equation}
with
\begin{align}
\overline{B}_3&(z_1,z_2) = \,\Ef{-1}{\infty}{1} + \eps \,\Big[\alpha_1\,\Ef{-1&1}{\infty&0}{1}+\alpha_2\,\Ef{-1&1}{\infty&1}{1}\\
\nonumber&\,+\alpha_3\,\Ef{-1&1}{\infty&\lambda_1}{1}+\alpha_4\,\Ef{-1&1}{\infty&\lambda_2}{1}\Big]
+\eps^2\,\Big[\alpha_1^2\,\Ef{-1&1&1}{\infty&0&0}{1}\\
\nonumber&\,
+\alpha_1\alpha_2\,\Big(\Ef{-1&1&1}{\infty&0&1}{1}+\Ef{-1&1&1}{\infty&1&0}{1}\Big)+\alpha_1\alpha_3\,\Big(\Ef{-1&1&1}{\infty&0&\lambda_1}{1}+\Ef{-1&1&1}{\infty&\lambda_1&0}{1}\Big)\\
\nonumber&\,+\alpha_1\alpha_4\,\Big(\Ef{-1&1&1}{\infty&0&\lambda_2}{1}+\Ef{-1&1&1}{\infty&\lambda_2&0}{1}\Big)+\alpha_2^2\,\Ef{-1&1&1}{\infty&1&1}{1}\\
\nonumber&\,
+\alpha_2\alpha_3\,\Big(\Ef{-1&1&1}{\infty&1&\lambda_1}{1}+\Ef{-1&1&1}{\infty&\lambda_1&1}{1}\Big)+\alpha_2\alpha_4\,\Big(\Ef{-1&1&1}{\infty&1&\lambda_2}{1}+\Ef{-1&1&1}{\infty&\lambda_2&1}{1}\Big)\\
\nonumber&\,
+\alpha_3^2\,\Ef{-1&1&1}{\infty&\lambda_1&\lambda_1}{1}+\alpha_3\alpha_4\,\Big(\Ef{-1&1&1}{\infty&\lambda_1&\lambda_2}{1}+\Ef{-1&1&1}{\infty&\lambda_2&\lambda_1}{1}\Big)\\
\nonumber&\,+\alpha_4^2\,\Ef{-1&1&1}{\infty&\lambda_2&\lambda_2}{1}\Big]+\ord(\eps^3)\,.
\end{align}
We observe again that all master integrals are uniform in weight order by order in the $\eps$ expansion.

%% file: conclusion.tex

\section{Conclusion}
\label{sec:conclusion}

In this paper we have introduced a class of iterated integrals on elliptic curves
that have at most logarithmic singularities, and therefore deserve to be called \emph{elliptic polylogarithms}.
The key idea is that the kernels that define the iterated integrals are obtained by analysing 
the abelian differentials of the first, second and third kinds on an elliptic curve, requiring the introduction of root-valued integration kernels. This idea by itself is not new, and iterated integrations over kernels involving square roots have been considered before, cf., e.g., refs.~\cite{Aglietti:2004tq,Bonciani:2010ms,Ablinger:2014bra}. The main difference between the functions in the literature and our elliptic polylogarithms is that we insist on having integration kernels with at most simple poles, leading to iterated integrals with at most logarithmic singularities on the elliptic curve. For this reason, we do not include abelian differentials of the second kind into our basis of integration kernels (which by definition have poles of higher order), and we are forced to consider integration kernels over transcendental functions, in particular incomplete elliptic integrals of the second kind. The closure of the algebra of integration kernels then forces us to consider two infinite towers of integration kernels for each point on the elliptic curve (one of the two towers is absent for the branch points).

A large part of our paper was devoted to studying some of the properties of elliptic polylogarithms. First, we have clarified how the iterated integrals of root-valued elliptic kernels are related to the multiple elliptic polylogarithms considered in the mathematics literature~\cite{BrownLevin,MatthesThesis,Broedel:2014vla}. We have shown that, under the isomorphism which identifies an elliptic curve with a torus, our integration kernels are mapped to simple linear combinations of the integration kernels introduced in refs.~\cite{BrownLevin,MatthesThesis,Broedel:2014vla} (up to the distinction that we prefer to work with meromorphic rather than periodic functions). Since the kernels of refs.~\cite{BrownLevin,MatthesThesis,Broedel:2014vla} are known to be independent, this proves at the same time the independence of our integration kernels. Second, we have provided an explicit algorithm to compute primitives of elliptic polylogarithms multiplied by rational functions. This algorithm generalises to elliptic polylogarithms the classical algorithm to compute primitives of rational functions on an elliptic curve in terms of the elliptic integrals of the first, second and third kinds. Finally, we have applied our results to compute the $\eps$-expansion of certain classes of ${}_2F_1$ and Appell $F_1$ function that cannot be expressed in terms of ordinary MPLs. We observe in all cases that it is possible to choose the master integrals in such a way that the results have uniform weight order by order in the $\eps$-expansion, hinting at an extension of the concept of ``pure function'' well-known from ordinary polylogarithms. We emphasise that in order for this to be true, it is important that the weight is \emph{not} identified with the number of integrations.

Our elliptic polylogarithms are very flexible and incorporate large classes of other special functions. We give here a list of functions which can be expressed in terms of our elliptic polylogarithms:
\begin{itemize}
\item Ordinary MPLs evaluated at arguments that are rational functions on the elliptic curve, cf.~eq.~\eqref{eq:E_to_G} and section~\ref{sec:G_elliptic}.
\item The (incomplete) elliptic integrals of the first, second and third kinds, cf.~eq.~\eqref{eq:E3_to_Pi} and~\eqref{eq:Z3_to_E}.
\item The elliptic polylogarithms of refs.~\cite{BrownLevin,MatthesThesis,Broedel:2014vla}.
\item It was shown~\cite{PanzerETH} that some instances of the $\textrm{ELi}$ functions of refs.~\cite{Adams:2014vja,Adams:2015gva,Adams:2015ydq} can be expressed in terms of the elliptic polylogarithms of refs.~\cite{BrownLevin,MatthesThesis,Broedel:2014vla}. As a consequence, these functions can also be written in terms of our elliptic polylogarithms.
\item The elliptic generalisations of polylogarithms of ref.~\cite{Remiddi:2017har} are a special case of the functions considered here.
\end{itemize}
Since many of these functions have appeared in physics computations, we foresee that the elliptic polylogarithms introduced in this paper will play a prominent role in applications to Feynman integrals. 
In a companion paper~\cite{plumber_paper}, we have applied our formalism, in particular the integration algorithm introduced in section~\ref{sec:algorithm}, to the computation of the sunrise integral with three different masses in two space-time dimensions. It will be exciting to see for which other Feynman integrals our elliptic polylogarithms provide the natural language.

We conclude this paper by commenting on some limitations and directions for future research. Throughout this paper, we have assumed that the branch points $a_i$ that define the elliptic curve are held constant. In applications, however, it is often the case that the branch points are themselves dynamical variables, and one wishes to differentiate and/or integrate with respect to the branch points. This situation is typical for Feynman integrals, where one usually invokes differential equations to obtain analytic results. The latter lead to iterated integrals in the branch points of the elliptic curve, or equivalently the modular parameter $\tau$ that defines the lattice, cf., e.g.,~\cite{Adams:2014vja,Adams:2015gva,Adams:2015ydq,Adams:2016xah,Ablinger:2017bjx,Remiddi:2016gno}. The functions introduced in this paper are not adapted to such a scenario. It is known, however, that there is a close relationship between the iterated integrals of refs.~\cite{BrownLevin,MatthesThesis,Broedel:2014vla} and iterated integrals in the modular parameter $\tau$, in particular iterated integrals over modular forms, cf.~refs.~\cite{Brown:mmv,Broedel:2015hia,MatthesThesis,Matthes:2016mrd,Matthes:Eisenstein2,Matthes:QuasiModular}. The latter are known to show up also in Feynman integral computations~\cite{Adams:2016xah}. It will be important to extend our framework to include differentiation and integration with respect to the branch points, and it will be fascinating to understand in detail the connection between the functions introduced in this paper and the theory of iterated integrals over modular forms.

%% file: regularisation.tex

\section{Regularisation}
\label{app:shuffle_regularisation}

In this section we discuss how to define a regularised version of (elliptic)
polylogarithms. Indeed, the recursive definitions in eq.~\eqref{eq:MPL_def},
\eqref{eq:eMPLs_def} and eq.~\eqref{eq:E4_def} may exhibit end-point
singularities at the origin $x=0$ whenever $c_k=0$. In the case of ordinary
MPLs, the regularisation is implemented through the special treatment of
polylogarithms where all $c_i$ are zero (see eq.~\eqref{eq:log_def}), and it
effectively amounts to a deformation of the integration contour (because the
integration starts from $x=1$ in eq.~\eqref{eq:log_def}, and not from $x=0$ as
in eq.~\eqref{eq:MPL_def}). While easy to implement in practise, this
regularisation procedure is rather subtle, because it is a priori not clear
that a deformation of the contour will preserve the shuffle algebra properties
of MPLs, see eq.~\eqref{eq:shuffle_G}.
Indeed, the derivation of eq.~\eqref{eq:shuffle_G} relies on the fact that both
factors in the left-hand side are iterated integrals over the \emph{same}
contour, and so it is not clear that the prescription in eq.~\eqref{eq:log_def}
preserves the shuffle product. In this appendix we review a general procedure
to regularise iterated integrals with logarithmic singularities in a way that
preserves the shuffle algebra structure, and we show that this procedure
reduces to the usual prescription for MPLs in eq.~\eqref{eq:log_def}.

\subsection{Shuffle regularisation}

Consider a function $f(x)$ that has at most logarithmic singularities. For every point $x_0\in\widehat{\mathbb{C}}$ there is a neighbourhood $U$ such that for every $x\in U$ we can write
\beq
f(x) = \sum_{k=0}^nf_k(x)\,\log^k(x-x_0)\,,
\eeq where $n$ is a positive integer and the functions $f_k$ are holomorphic on $U$. The \emph{regularised value of $f$ at $x_0$} is defined by (see, e.g., ref.~\cite{Brown:2008um})
\beq
\textrm{Reg}_{x_0}f(x) \equiv f_0(x)\,.
\eeq
The map $\textrm{Reg}_{x_0}$ has the following properties:
\begin{enumerate}
\item If $f$ has at most logarithmic singularities, then $f_0$ is regular at $x=x_0$.
\item If $f$ is regular at $x=x_0$, then $\textrm{Reg}_{x_0}f(x) = f(x)$, i.e., the regularised value of a regular quantity does not change.
\item The map $\textrm{Reg}_{x_0}$ is an algebra homomorphism, i.e., it preserves the multiplication,
\beq\label{eq:shuffle_homo}
\textrm{Reg}_{x_0}(f(x)\,g(x)) = \textrm{Reg}_{x_0}f(x)\,\textrm{Reg}_{x_0}g(x)\,.
\eeq
\end{enumerate}
A natural way to define a regularised version of polylogarithms is to replace all functions by their values regularised at the origin,
\beq
G(\vec c;x)\to \textrm{Reg}_0G(\vec c;x) {\rm~~and~~} \textrm{E}_N\!\left(\begin{smallmatrix}\vec n\\\vec c\end{smallmatrix};x\right) \to \textrm{Reg}_0\textrm{E}_N\!\left(\begin{smallmatrix}\vec n\\\vec c\end{smallmatrix};x\right)\,,\quad N=3,4\,.
\eeq
It is usually assumed in the literature that one is talking about the regularised version, and and the map $\textrm{Reg}_0$ is usually not written explicitly.
Equation~\eqref{eq:shuffle_homo} ensures that if we replace all polylogarithms by their values regularised at the origin, the shuffle-algebra structure is preserved. 

While the previous discussion allows one to define a regularised version of iterated integrals that is compatible with the shuffle algebra structure, it is less well suited for explicit computations, because the regularisation procedure requires one to first work with an arbitrary base-point $\delta$ and then to expand around $\delta=0$ and to remove the logarithmic divergences.
In the remainder of this section we show that for ordinary MPLs the effect of the map $\textrm{Reg}_0$ is equivalent to the well-known prescription in eq.~\eqref{eq:log_def}, and we then propose an extension of this prescription to elliptic polylogarithms.

\subsection{Regularisation of ordinary MPLs}
It is easy to see that the integral in eq.~\eqref{eq:MPL_def} is divergent precisely when $c_k=0$. Using the shuffle algebra, we can always write any MPL in as a linear combination of terms involving MPLs where the last index is not zero, except for terms where all indices are zero. For example, we have
\beq
G(1,0,0;x) = G(0,0;x)\,G(1;x) - G(0;x)\,G(0,1;x) + G(0,0,1;x)\,.
\eeq
We therefore only need to consider the regularisation of MPLs where all indices are zero. Let us first compute the regularised version of $G(0;x)$. We find
\beq
\textrm{Reg}_0G(0;x) = \textrm{Reg}_0\int_{\delta}^x\frac{dt}{t} = \textrm{Reg}_0\left(\log x-\log\delta\right) = \log x\,.
\eeq
We see that the shuffle-regularised version of $G(0;x)$ is precisely given by the prescription in eq.~\eqref{eq:log_def}. Since the map $\textrm{Reg}_0$ preserves the shuffle algebra structure, this result immediately extends to higher weights,
\beq
\textrm{Reg}_0G(\underbrace{0,\ldots,0}_{k\textrm{ times}};x) = \textrm{Reg}_0\frac{1}{k!}G(0;x)^k = \frac{1}{k!}\left[\textrm{Reg}_0G(0;x)\right]^k = \frac{1}{k!}\log^kx\,.
\eeq
To conclude, we see that the regularisation at $x=0$ is completely equivalent to the prescription in eq.~\eqref{eq:log_def} commonly used in the physics literature.

\subsection{Regularisation of elliptic polylogarithms}
The equivalence of the regularisation using the map $\textrm{Reg}_0$ and the prescription in eq.~\eqref{eq:log_def} in the case of ordinary MPLs is particularly convenient in computations. In the case of elliptic polylogarithms, however, we were unable to find such a simple contour prescription to implement the map $\textrm{Reg}_0$. The main obstacle is the fact that we need to consider several integration kernels that all have a logarithmic singularity at the origin. For example, while $\textrm{Reg}_0\Et{1}{0}{x} = G(0;x) = \log x$, we have
\beq
\textrm{Reg}_0\Et{-1}{0}{x} = \textrm{Reg}_0\int_{\varepsilon}^x\frac{y_0\,dx}{xy} = \log x + \int_{0}^x{dx}\left(\frac{y_0}{xy} -\frac{1}{x}\right)\,.
\eeq
We are unaware of a simple prescription to implement this regularisation by deforming the integration contour.

While it is of course possible to work directly with the map $\textrm{Reg}_0$, we find it more convenient to redefine the basis of integration kernels in such a way that only $\varphi_1(0,x)$ has a pole at $x=0$. More precisely, for $n\notin\{0,1\}$, we define
\beq\bsp\label{eq:phi_reg_def}
\varphi_{n}^{\textrm{reg}}(c,x)&\, \equiv \varphi_{n}(c,x) -{\delta_{c0}}\,Z_3^{(n-1)}(0)\,\varphi_1(0,x) \,,
\esp
\eeq
and the kernels $\varphi_0$ and $\varphi_1$ remain unchanged,
\beq
\varphi_{0}^{\textrm{reg}} \equiv \varphi_0 {\rm~~and~~}\varphi_{1}^{\textrm{reg}} \equiv \varphi_1\,.
\eeq
 If we use the $\varphi_{n}^{\textrm{reg}}$ to define the elliptic polylogarithms $\textrm{E}_3$ in eq.~\eqref{eq:eMPLs_def}, then we see that the iterated integral in eq.~\eqref{eq:eMPLs_def} converges unless $(n_k,c_k)=(1,0)$. This latter case is identical to the case of ordinary MPLs, and so we define
\beq
\textrm{E}_3\!\big(\underbrace{\begin{smallmatrix}1&\ldots&1\\ 0&\ldots&0\end{smallmatrix}}_{k \textrm{ times}};x\big) \equiv G(\underbrace{0,\ldots,0}_{k\textrm{ times}};x) = \frac{1}{k!}\log^kx\,.
\eeq
This prescription is very simple to implement in practise, and it preserves the shuffle algebra structure because it coincides with the map $\textrm{Reg}_0$. 

The previous discussion easily extends to the quartic case by defining, for $n\notin\{0,1\}$,
\beq\label{eq:psi_reg_def}
\psi_{\pm n}^{\textrm{reg}}(c,x)  \equiv \psi_{\pm n}(c,x) -{\delta_{c0}}\,Z_4^{(n-1)}(0)\,\psi_1(0,x) \,,
\eeq
and the integration kernels for $n\in\{0,1\}$ remain again unchanged. 

\subsection{Regularised special values}
Regularisation also plays an important role when considering values of elliptic polylogarithms at certain special points at which they may diverge. It often happens in applications that individual terms diverge logarithmically, but the divergences cancel in the sum. There may, however, be finite terms left over that need to taken into account correctly. An example of this is given in section~\ref{sec:examples}, cf. eq.~\eqref{eq:reg_values} and eq.~\eqref{eq:reg_values_2}.

More precisely, $\textrm{E}_N\!\left(\begin{smallmatrix}n_1&\ldots& n_k\\ c_1&\ldots& c_k \end{smallmatrix};x\right)$, $N=3,4$, may diverge logarithmically whenever $x$ approaches $c_1$. In the following we show how we can extract the logarithmic singularities. We first apply some results on shuffle algebras from ref.~\cite{Panzer:2014caa} (in particular Definition 2.1 and Lemma 2.2), which allow us to rewrite $\textrm{E}_N\!\left(\begin{smallmatrix}n_1&\ldots& n_k\\ c_1&\ldots& c_k \end{smallmatrix};x\right)$ as a linear combination where the only terms that diverge as $x\to c_1$ have the form $\textrm{E}_N\!\left(\begin{smallmatrix}n_1&\ldots& n_k\\ c_1&\ldots& c_1 \end{smallmatrix};x\right)$. In each of these terms we replace the integration kernels by a regularised version very similar to eq.~\eqref{eq:phi_reg_def} and eq.~\eqref{eq:psi_reg_def} for $n> 1$ or $n\le -1$. For example in the cubic case we apply the identity
\beq\bsp\label{eq:phi_reg_def}
\varphi_{\pm n}(c_1,x)&\, = \varphi_{\pm n}^{\textrm{reg}}(c_1,x) +Z_3^{(n-1)}(c_1)\,\varphi_1(c_1,x) \,.
\esp
\eeq
All logarithmic singularities are now explicit, and divergences only arise from the integration kernel $\varphi_1(c_1,x)$. We can again use the shuffle algebra properties to isolate all divergent terms in factors of the form $\textrm{E}_N\!\left(\begin{smallmatrix}1&\ldots& 1\\ c_1&\ldots& c_1 \end{smallmatrix};x\right)$. The remaining integrals are finite, and we define, for $n_1\neq 1$,
\beq
\ERegt{n_1&\ldots &n_k}{c_1&\ldots& c_1} \equiv \int_0^{c_1}dt_1\,\varphi_{ n_1}^{\textrm{reg}}(c_1,t_1)\int_0^{t_1}\ldots \int_0^{t_{k-1}}dt_k\,\varphi_{ n_k}^{\textrm{reg}}(c_k,t_k)\,.
\eeq
In this way we recover immediately the results in eq.~\eqref{eq:reg_values} and eq.~\eqref{eq:reg_values_2}.
The extension to the quartic case is straightforward.